\documentclass[nofootinbib,11pt,showkeys]{article}
\usepackage[utf8]{inputenc}
\usepackage[]{graphicx}
\usepackage{amsmath}
\usepackage{mathptmx}
\usepackage{amssymb}
\usepackage{color}
\usepackage{todonotes}
\usepackage[margin=1in]{geometry}
\usepackage{booktabs}
\usepackage{enumerate}
\usepackage{url}
\usepackage[export]{adjustbox}[2011/08/13]
\usepackage[figuresleft]{rotating}

\usepackage[round, sort]{natbib}

\usepackage[T1]{fontenc}
\usepackage{setspace}
\usepackage[activate={true,nocompatibility},final,kerning=true,spacing=true,factor=1100,stretch=10,shrink=10]{microtype}

\usepackage[lining]{ebgaramond}
\usepackage[cmintegrals,cmbraces]{newtxmath}
\usepackage{ebgaramond-maths}
\usepackage{relsize}
\let\oldtexttt\texttt
\renewcommand{\texttt}[1]{\oldtexttt{\small #1}}

\usepackage{caption}
\usepackage{subcaption}
\DeclareCaptionFormat{myformat}{#1#2#3\dotfill}
\usepackage{parskip}
\usepackage[hang,flushmargin]{footmisc} 

\usepackage{url}
\usepackage{hyperref}
\hypersetup{
     colorlinks=true,
     linkcolor=black,
     filecolor=black,
     citecolor = black,      
     urlcolor=black}

\renewcommand{\thepage}{\footnotesize\arabic{page}}   

\begin{document}

\title{\textbf{A political cartography of news sharing: Capturing story, outlet and content level of news circulation on Twitter}}
\date{}
{\Large\textbf{A political cartography of news sharing:\\[1mm] Capturing story, outlet and content level of news circulation on Twitter}}\\[5mm]
Felix Gaisbauer$^{1}$, 
Armin Pournaki$^{2,3,4}$, 
Jakob Ohme$^{1}$\\[4mm]
{\small
\noindent $^1$ Weizenbaum Institute for the Networked Society, Berlin, Germany \\ $^2$ Max Planck Institute for Mathematics in the Sciences, Leipzig, Germany \\ 
$^3$ Laboratoire Lattice, CNRS \& ENS-PSL \& Université Sorbonne nouvelle, Paris, France \\ $^4$ Sciences Po, médialab, Paris, France \\[1mm]
Corresponding author: Felix Gaisbauer, \texttt{felix.gaisbauer@weizenbaum-institut.de}}\\[5mm]
\pagenumbering{arabic}
\renewcommand{\thepage}{\footnotesize\arabic{page}}
{\small
\textbf{Abstract}\\
\noindent News sharing on digital platforms shapes the digital spaces millions of users navigate. Trace data from these platforms also enables researchers to study online news circulation. In this context, research on the types of news shared by users of differential political leaning has received considerable attention. We argue that most existing approaches (i) rely on an overly simplified measurement of political leaning, (ii) consider only the outlet level in their analyses, and/or (iii) study news circulation among partisans by making ex-ante distinctions between partisan and non-partisan news. In this methodological contribution, we introduce a research pipeline that allows a systematic mapping of news sharing both with respect to source and content. As a proof of concept, we demonstrate insights that otherwise remain unnoticed: Diversification of news sharing along the second political dimension; topic-dependent sharing of outlets; some outlets catering different items to different audiences.
\\[3mm]
}
\section{Introduction}
\label{intro}
News sharing on digital platforms is a crucial activity that shapes the digital spaces millions of users navigate. With an ever-increasing number of users encountering news content via social media, sharing a news article can influence the visibility of the item in global networks as well as in small networked publics. News outlets rely more and more on their content being circulated online -- and receive almost instant feedback on the impact of their stories, which in turn affects editorial decisions. Especially for alternative news outlets, digital platforms are the key to success. Little financial resources are needed to publish and distribute news there, and social media allows them to bypass pre-digital media systems' critique and verification mechanisms. Their reach in these spaces critically depends on the sharing behavior of their followers. 

The abundance of information we can encounter online shifts ``the function of curating content from newsroom editorial boards to individuals, their social networks, and manual or algorithmic information sorting'' \cite[p. 1130]{bakshy_exposure_2015}. 
Consequently, much attention has been devoted to the study of news consumption and circulation -- especially among political groups, partisans or politically interested users (see, for example, \cite{chadwick_news_2019,eady_news_2020,wischnewski_shareworthiness_2021,osmundsen_partisan_2021,guess_cracking_2021}). 

Yet, these analyses of web link data -- the ``near-universal currency of social media information exchange'' \cite[p. 6]{eady_news_2020} -- are usually based on a rather restricted set of news features. Mostly, they focus on which outlets the news come from and where the users sharing them are located on a one-dimensional ideological axis. This underestimates the complexity of the phenomenon. On the one hand, a one-dimensional political space might be too simplified for countries with other political settings than a two-party system. On the other, research on why users share news has shown that the act of sharing a news item combines attitudinal, informational and social aspects. Users might, for example, want to display their opinion, recommend valuable information, make social connections to others, or present sources they identify with \citep{kumpel_news_2015,trilling_drivers_2023,green_curation_2021,weeks_incidental_2017}. Accordingly, different aspects of a news item -- its source, its topic, its ideological congruency, etc. -- can all be of elevated or of marginal importance, depending on the circumstances (see e.g. \cite{trilling_newsworthiness_2017}). The information environment within networked publics is shaped by a combination of these factors. To mirror this complexity, more sophisticated methodological approaches are needed.

In light of these considerations, it appears imperative to obtain a systematic mapping of the social media information ecosystem. Especially with respect to news circulation among political groups and partisans, advances towards a mapping that do not focus on a limited subset of features of news (such as their outlet, or whether they are false or politically extreme) exist, but these efforts have remained of insular nature.
This article aims to develop such a mapping in a holistic way. Its contribution is, first and foremost, \textit{methodological}. It introduces a research pipeline that produces a \textit{political cartography of news sharing}: News sharing is mapped in a multi-dimensional political space and can be studied not only regarding news sources, but also news content. 

More precisely, the introduction of the research pipeline is divided into three parts. \textit{First}, we draw on previous theoretical and empirical work on political spaces, news sharing and news effects to point out three methodological shortcomings that have impeded a more comprehensive overview of news sharing on social media (Sec. \ref{method_concerns}). Efforts to overcome these shortcomings have remained insular in nature. 
Hence, \textit{secondly}, we introduce the research pipeline in detail which enables us to bypass these shortcomings and study political news sharing on Twitter in a more comprehensive way (Sec. 3). The pipeline combines political space construction via network embeddings, news link sharing data from Twitter and semi-automated text analysis of the corresponding article full texts. The novelty of the approach lies in the combination of state-of-the-art network embeddings with multi-scale content- and outlet-level analyses. \textit{Thirdly}, we provide a proof of concept: Findings derived from this political cartography of news sharing that would have been overlooked in most previous study designs (Sec. 4). Based on a news sharing data set from the German Twittersphere of March 2023, we show that crucial patterns in the two-dimensional political dispersion of news -- either theoretically predicted or empirically pointed to by research mentioned in Sec. 2 -- emerge on the \textit{story}, \textit{outlet}, and \textit{topic} level.\footnote{We stress here that this study is concerned with the introduction of the research pipeline and findings that show its value. A comprehensive analysis of the systemic patterns of news sharing is out of the scope of this contribution, but will be provided in separate work.} Independent of the platform under investigation, the study calls for multi-level and -dimensional studying of news sharing.
%
%
%
%
\section{Three methodological concerns} 
\label{method_concerns}
\subsection{The prevalence of one-dimensional measures of ideology}
Which types of news are shared by political partisans? Who of them shares false news? How do content and frequeny of news engagement differ across the political spectrum? To answer such questions, a proxy of users' political leaning is needed. Such a proxy is very often provided by a unidimensional representation of ideology -- typically, a classical left-right axis. User behavior, such as which news outlets these users tend to engage with, is mapped along this axis. Ideology is measured either directly via surveys or users' online self-representation \citep{wischnewski_shareworthiness_2021,bakshy_exposure_2015,osmundsen_partisan_2021}, or with embedding methods based on social media interaction networks \citep{barbera2015birds,bond2015quantifying}. 

Survey approaches allow measurement of several ideological dimensions, but are resource-intensive especially if they are to be combined with digital trace data \citep{wojcieszak_no_2023,guess_cracking_2021,osmundsen_partisan_2021,ohme2024digital}. On the other hand, using users' online self-identification, for example identifying markers like flags or emojis, allows to capture only political dichotomies instead of continuous spectra.
This is why embedding approaches have enjoyed sustained popularity in recent years (see e.g. \cite{barbera2015tweeting,ramaciotti_morales_inferring_2022,del_pozo_hidden_2023,gaisbauer2023grounding,barbera2015birds,bond2015quantifying}). While they can, in principle, be used to infer an arbitrary number of political dimensions, these methods were mostly developed to study online public debate in the United States. Therefore, they are historically tied to a two-party system for which a one-dimensional space might be suitable, and for which the axis has a straightforward interpretation (Democrat-leaning to Republican-leaning, Liberal-leaning to Conservative-leaning, etc.). Ready-to-use software packages that were published along with said approaches, such as in \cite{barbera_gitideology}, could be used by other researchers -- but then they were bound to stick to the one-dimensional measure unless they were willing and able to modify the software. For these reasons, studies of other countries often adopt the unidimensional political space \citep{aruguete_news_2021,del_pozo_hidden_2023}. Still, in different political circumstances (such as in Germany), a one-dimensional political space often is too simplified. Political science has long suggested that, for example, for the majority of European contries -- including Germany --, a two-dimensional political space is needed to reflect major political fault lines in a satisfactory way \citep{wheatley2021reconceptualizing,kriesi2006globalization,olbrich2021rise}. This appears to hold both on the supply side \citep{kriesi2006globalization} (i.e., with respect to the profile of political parties), as well as on the demand side \citep{wheatley2021reconceptualizing} (i.e., voter attitudes). Germany poses a suitable example to employ a two-dimensional political space because right-wing populist movements that have established themselves in many European countries (such as France) emerged comparatively late. It hence poses a good minimal expample for finding pronounced differences in news sharing between right-wing populist partisans, that differ in several respects from traditional conservative parties, and others.

As has been stated, network embedding methods do allow for an arbitrary number of dimensions. Even early works on latent space models \citep{hoff2002latent}, on which recent approaches are based, inferred latent spaces with more than one dimension. The obvious challenge of more complex political spaces is the interpretation of the axes: Which dimension (or combination of dimensions) corresponds to specific political fault lines is not immediately clear. Recently, scholars have developed approaches in which low-dimensional ideological spaces inferred from digital trace data are enriched with data from expert surveys, such that the axes can be interpreted as corresponding to certain political issues measured in these surveys \citep{ramaciotti_morales_inferring_2022}. We will follow a similar approach to arrive at a two-dimensional political space, which is complex enough to account for Germany's major political divisions, but at the same time simple enough to prevent diluting systematic findings with too much detail. 

Whether a two-dimensional political space reveals systematic insights -- that is, meso- or macro-scale patterns in the newly-constructed two-dimensional spce -- about news sharing that remain obstructed in the one-dimensional view is an empirical question. Our approach is useful if it can yield such insights.
\newpage
Hence, we hypothesize a proof of concept:
\begin{quote}
	H1: A two-dimensional political space reveals systematic insights that are unobservable in a one-dimensional (left-right) approach.
\end{quote}

\subsection{Restriction to the outlet level}

Analyses of news sharing based on digital trace data \citep{eady_how_2019,guess_almost_2021,cinelli_selective_2020} have the advantage of high ecological validity: They can observe actual behavior of potentially millions of users. But despite the wealth of data these studies investigate, they mostly focus on analyses on the outlet level.\footnote{This similarly holds for audience research in general, see e.g. \cite{mukerjee_networks_2018,garimella_political_2021}.} Individual news items, events, or topics are all subsumed under the outlet that publishes them. Outlets can then be located on the ideological spectrum based on the users that share them, or the overlap of news (sharing) diets of different user groups can be calculated. Often, studies found surprisingly high outlet overlap, which appeared as a conundrum in light of increasingly polarized debate on- and offline \citep{post_issue_2023}. 

This conundrum might be explained relatively easily: Newsfeed environments contribute to the `unbundling' of news \citep{trilling_conceptualizing_2019}, that is, they disconnect the news item from its source. News stories are not read any more (if they ever were) in a pre-fabricated combination with other items, such as in a physical newspaper that covers and might juxtapose different viewpoints within one publication. Users come across isolated news stories from a variety of outlets in digital spaces -- both incidentally and selectively. Outlet level-only approaches cannot account for this development. Analysing news sharing \textit{solely} with respect to the outlets is justified the outlet remains the decisive factor in sharing decisions even in unbundled environments.

A few larger-scale studies attempt to go beyond the outlet level, such as \cite{wojcieszak_no_2023} (political vs. non-political news), \cite{del_pozo_hidden_2023} (sentiment towards politicians), or \cite{trilling_newsworthiness_2017} (message characteristics of news articles).
Notably, \cite{green_curation_2021} have uncovered significantly more political segregation in news sharing in the US if the individual story was considered instead of the outlet level. For outlets usually referred to as `moderate' or `non-partisan', that is, for which the average political leaning of a user sharing the outlet was between Democrats and Republicans, individual articles were often either shared mostly/only by Republicans or mostly/only by Democrats.\footnote{Generally, it might also be advisable to not only look at the average political position of a user sharing a news source, but their distribution, as we will see below.} 

Content-wise, users might share only specific news items while choosing them from a wide range of outlets. Previous research also points in this direction: \cite{von_nordheim_im_2020} study news sharing by politicial elites -- German MPs -- and find that source diversity does not necessarily translate into content diversity. \textit{AfD} politicians shared links from many different news outlets, but focused on a comparatively narrow set of topics. Whether the same holds for regular users, too, is yet to be determined.

Both of the above angles of analysis -- considering individual stories or topics covered -- might provide, as has been indicated by previous research, important additions to the outlet-centered view. Yet, we cannot expect that one of the two angles suffices to explain all heterogeneity in news sharing. Users across the political spectrum might share articles from a certain outlet on a certain topic; but that different partisans still do not share the same articles -- because, say, they share opinion pieces on the topic that voice detrimental opinions -- can then only become apparent in the story-level view. On the other hand, if we find that different partisans largely share different news stories from an outlet, the metatopic level can show whether this is due to a focus on different content. This is why it is important to include both levels of analysis.
We hence additionally hypothesize for our proposed research pipeline: 

\begin{quote}
	H2a: Methodological inclusion of the news story level reveals systematic insights that are unobservable in a purely outlet-centered approach.
	
	H2b: Methodological inclusion of the topic level reveals systematic insights that are unobservable in a purely outlet-centered approach.
\end{quote}



\subsection{Top-down approaches}

As has been mentioned above, there already is some research that does not only investigate news sharing only with respect to the outlets that are shared. \cite{wojcieszak_no_2023} differentiate between political and non-political news, for example. Yet, finer-grained distinctions in terms of content or ideological slant of news -- that are necessary to arrive at more specific findings -- are most often determined \textit{ex ante} and with respect to clearly `problematic' news. This means that either outlets altogether, or individual news pieces, were marked as for example misinforming \citep{osmundsen_partisan_2021,vosoughi2018spread} or highly partisan \citep{wischnewski_shareworthiness_2021}. 

Such approaches are undoubtedly useful in many study attempts, but also reduce the complexity of news sharing into prefabricated, binary categories. Whether information circulating online contributes or leads to biased opinions is still an open question, but the impact of both strongly partisan news \citep{guess_consequences_2021} and misinformation \citep{altay2023misinformation} is questioned by recent research. The effects of these types of news on opinions have been found to be comparably small, such that it might be advisable to ``move away from fake news and blatantly false information as their prevalence and effects are likely minimal. Instead, we should investigate subtler forms of misinformation that produce biased perceptions of reality'' \cite[p. 9]{altay2023misinformation}. Such `subtler forms of misinformation' might not consist of individually misinforming news pieces at all. Rather, the \textit{sum} of news pieces circulating among certain political partisans might systematically exclude certain views or topics, and focus very strongly on others. Behavior-based, inductive approaches can uncover emergent systemic patterns in news sharing that might otherwise be overlooked. This might be especially fruitful if these approaches operate at different levels of granularity at the same time. Then, a comprehensive picture of news sharing can be supplied.\footnote{This comprehensive picture can of coures \textit{lead to} hypotheses that might be more suitable for research making emprically informed ex-ante distinctions based on these results.} 
Content-wise, while everybody might be sharing news about a certain larger issue (the Russian aggression against Ukraine, for example), different groups might focus on different aspects of it (whether it makes sense to deliver tanks to Ukraine, or to what extent the EU sanctions against Russia create economic harm in Europe, for example). The research pipeline we develop allows a distinction between these levels in a bottom-up approach. Hence, we hypothesize: 

\begin{quote}
H3: Methodological inclusion of content at different levels of granularity reveals systematic differences between different political partisans.
\end{quote}
Answers to H1-H3 serve as indicators whether the pipeline we introduce in the following has the potential to improve research on news sharing. If we can answer them positively, they serve as `proofs of concept'.

\section{Method and research pipeline}

The methodological advance of this contribution lies in the innovative combination of multiple data sources and techniques of analysis. We use link sharing events, the follower network of German Members of Parliament (MPs), and article full texts via state-of-the-art network embedding methods and (semi-)automated text analysis. This enables the inference a multidimensional political space independently of the actual act of news sharing. First, we sketch out the inference of the political space from MP follower networks and its subsequent validation with party rankings from the CHES. Then, we outline the news link and corresponding full text collection, as well as the semi-automated approach to topic modelling. All in all, this pipeline allows us to politically map news sharing both by \textit{source} and by \textit{content} in a multidimensional political space.

\paragraph*{Sample}
Since the sample consists of three different yet interrelated data collections, we briefly sum up what data the pipeline relies on: 
\begin{enumerate}
	\item The Twitter follower network of (in this case 433 German) MPs, collected with the Twitter API v1.\footnote{For follower collection, the outdated v1 API access provided better performance than the Research API.} We only included users following at least three MPs in our analysis, which resulted in 1.2 million followers being included in our pipeline.
	\item All tweets from March 2023 including a link to at least one of the 26 outlets (see Appendix). Collection was carried out with the Twitter Research API at the beginning of April 2023 and resulted in roughly 500,000 unique tweets. The data set also includes shortened links (such as \texttt{bit.ly/AAAAA}). Sharing events can only be included in the analysis if the users sharing the news also appear in the embedding, i.e. follow at least three MPs. Coverage was 60 percent.
	\item The full texts of all articles that were not paywalled. 75 percent of all shares could be assigned a text (60,000 unique articles). Text collection was carried out with the \texttt{newspaper3k} Python library \citep{n3k-ouyang}. 
\end{enumerate}



\subsection{Inference of the political space}

The pipeline fundamentally rests on an estimation of users' political leanings. Such an estimation can be achieved by considering which politicians and political parties the users interact with on social media, for example by following them on Twitter. To this end, we collected the follower network of 433 MPs active on Twitter (list of MPs and inclusion criteria in Appendix). This follower network was used to estimate users' political positions in a lower-dimensional political space. We will first provide a high-level overview on the underlying assumptions and approach. This allows also researchers unfamiliar with network embeddings to get an overview of the method. 

\paragraph*{Overview} The fundamental assumption at play here in the estimation of political leaning is ideological \textit{homophily} \citep{mcpherson2001birds}: We assume that users tend to follow politicians that they agree with politically, that is, are ideologically \textit{close to}. This assumption has strong backing by previous research on social media follower networks \citep{himelboim2013birds,barbera2015birds,conover2012partisan}. In most cases, it replicates existent conventional, but more resource-intensive measures for ideology exceptionally well. For example, ideology estimates of users based on which politicians they followed showed very high agreement with voter registrations and campaign contributions in the US \citep{barbera2015birds}. \cite{himelboim2013birds} found that clusters of users following each other largely posted ideologically congruent content. These studies both worked with Twitter data, \cite{bond2015quantifying} provide similar results for users' political page connections on Facebook (again for the US). \cite{ramaciotti_morales_inferring_2022} find high agreement of follower network embeddings with text-based classifications of user self-descriptions for several ideological dimensions (France, Twitter).

Now, if each following decision can be seen as a noisy signal of ideological proximity, that is, a multi-dimensional vector of revealed preferences similar to a series of survey questions, one can infer a low-dimensional space with techniques of dimensionality reduction. We use Correspondence Analysis (CA) since we are faced with categorical data. The number of dimensions in the political space should balance comprehensiveness with parsimony. As elaborated above, political scholars suggest a two-dimensional space as the best fit for this necessity. Therefore, we use two Principal Components to construct a two-dimensional space, in which both regular users and MPs get assigned a position. Users/MPs with a similar following/follower profile get placed close to each other.

\begin{figure}[!t]
    \includegraphics[width = 19.5cm,center]{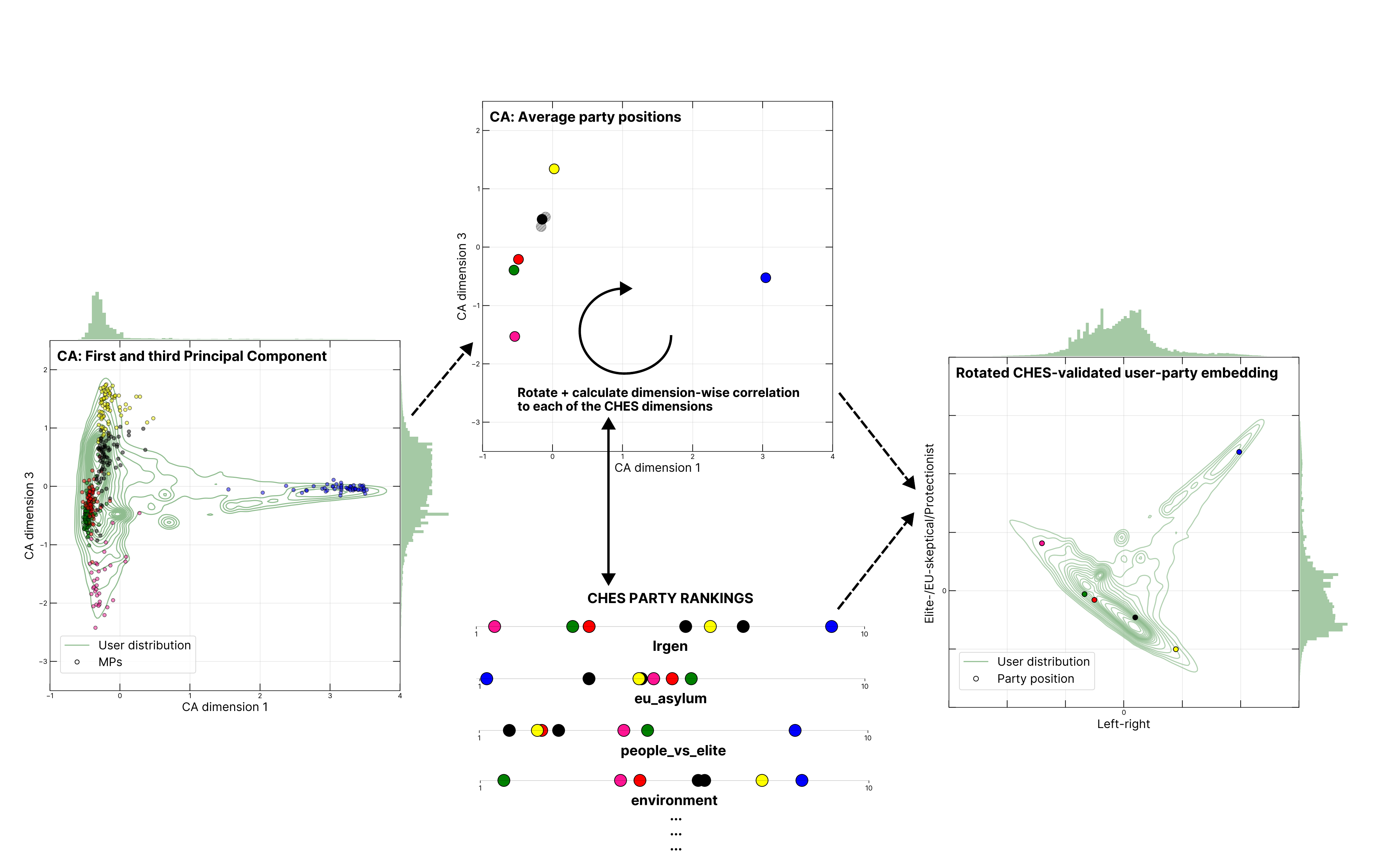}
    \caption{The process of identifying interpretable political axes from a two-dimensional space derived from Correspondence Analysis (CA). Left panel: Initial two-dimensional space that includes MPs represented by colored dots. Their colors are corresponding to their political party (i.e., \textit{AfD} blue, \textit{Die Linke} pink, \textit{Greens} green, \textit{CDU/CSU} black, \textit{FDP} yellow, \textit{SPD} red). The green density plot shows user distribution across this space, dimension-wise distributions are given on the margins. Middle: The average positions of each party's MPs are calculated (top). Subsequently, the space is rotated and the dimension-wise correlation with every othogonal combination of the left-right plus any other CHES dimension (examples displayed at bottom) is calculated, until we arrive at an angle for which two CHES dimensions maximize combined correlation. (\textit{CDU/CSU} are treated as separate parties to calculate the correlation, hence the two grey dots in the figure of the average party positions -- they represent their respective positions.) Then, the whole space (that is, including the non-MP users) is assigned the rotation and used as the basis of analysis in the following (right panel). (In fact, four CHES dimensions rank nearly equally high in orthogonal combination with the left- right axis with similar rotation angles. Hence we average over their four angles and combine them in the second dimension, which we call `Elite-/EU-skeptical/Protectionist'.)
    }
    \label{fig:ca-ches}
\end{figure}

At this stage, the task remains to assign meaning to directions in this space. For the simple one-dimensional case (which is, as we have sketched out above, done in most studies using embedding methods), its interpretation is easy. The most extreme politicians/parties on the spectrum and their respective political stances are used to give meaning to its directions. This usually results in it being interpreted as some kind of left-right axis. Considering more than one dimension brings more difficulty: Which path in the space now exactly represents, for example, a left-right dimension and which one corresponds to a different ideological fault line is more difficult to say. Based on a valuable procedure developed by \cite{ramaciotti_morales_inferring_2022}, we will use expert-provided national party positioning on ideology and policy issues for Germany from the Chapel Hill Expert Survey (CHES, \cite{jolly2022chapel}). The CHES is a survey administered among 421 political scientists. They provide positions of parties -- on either 7- or 10-point scales -- on ideological and policy issues. Such issues are, for example, `deregulation', `protectionism', or the general position of the party on a classical left-right scale. We introduce party positions into our embedding by calculating  the average position of the respective party's MPs. We then aim to find two directions in the space that are \textit{independent} of each other because then we can describe directions in the space without being redundant. Geometrically, this means that these directions are \textit{orthogonal dimensions}: That is, a change along one dimension does not affect the position in the other one. To ensure continuity with previous research using a left-right axis, we orthogonally combine the CHES left-right ranking (as an x-axis) with each other CHES issue position (as a y-axis). We then rotate the embedding degree-wise and calculate correlations with the two dimensions. Once we reach a maximum in the sum of correlations along the x- and y-axis, this pair of CHES issues helps us interpret x- and y-axis (see Fig. \ref{fig:ca-ches}). If several CHES issue dimensions have similarly high correlations for similar rotations, the second dimension is interpreted as their combination. (We could also just use the CHES issue with the \textit{highest} correlation to explain the dimension. But the other CHES issues will exhibit high correlation with the dimension even if we ignore them; including them hence can give the axis more explanatory power. This is especially useful considering that CHES asks for both high-level ideological dimensions (e.g., left-right) and more specific issues (anti-islam rhetoric). A combination of the latter gives the interpretation of the dimension more breadth, which is desirable if we operate in a two-dimensional space.)

\paragraph*{Procedure} To embed the followers of the MPs and the MPs themselves in a lower-dimensional space, we use CA \citep{bonica_mapping_2014}. The adjacency matrix $A$ of the user-MP follower network consists of columns representing MPs and rows representing users, and an entry is 1 if the user follows the respective MP, and 0 otherwise. This matrix is taken as a representation of a set of points in a multidimensional space. CA then produces a lower-dimensional space spanned up by its Principal Components via the calculation of the matrix of standardized residuals $S$ from $A$ and subsequent Singular Value Decomposition (SVD, for the mathematical details, see Appendix). Using CA \citep{barbera2015tweeting,ramaciotti_morales_inferring_2022,del_pozo_hidden_2023,bonica_mapping_2014} is computationally more efficient than using latent space models \citep{barbera2015birds,hoff2002latent,gaisbauer2023grounding}.\footnote{Even though both approaches are often employed in computational social research and CA has been validated \textit{with} latent space models as a baseline.} In the present case, with more than one million users being embedded, we hence choose CA. The closer a user and an MP are placed together in the space produced by CA, the higher the probability that the user follows the MP tends to be. We interpret the space as a political space and users placed close together as users sharing a political position. 

The CA-inferred two-dimensional political space is visible in the leftmost plot of Fig. \ref{fig:ca-ches}. How users are distributed in the space is indicated by the green density plot on top -- we also provide their dimension-wise distribution at the plot margins. All in all, we embedded nearly 1.2 million users (which is why we provide a density plot instead of their individual positions) and 433 MPs. 

We note here that we leave out the second Principal Component (PC) of the CA and operate with PCs 1 and 3. Along the second PC, MPs of the different parties are spread out far more widely than along the other axes (for a graphical display, see Appendix). Along dimensions 1 and 3, the MP positions of the different parties never have a variance larger than 0.2, while \textit{all} except one party has a variance larger than 0.2 along dimension 2. We find that the dimension correlates with the MPs' follower counts (Spearman Rank 0.79, Pearson 0.58) and thus rather reflects the MPs' popularity.\footnote{While it has been claimed that the normalization step in the CA amounts to something similar to the popularity parameters in latent space models \citep{bonica_mapping_2014}, CA apparently does at least not adequately factor out popularity of the MPs in this case.} We hence leave out the second dimension because it is not politically meaningful.

For each political party in parliament, we take its position as the average of the positions of its MPs (Fig. \ref{fig:ca-ches}, top middle). We then rotate the Principal Components of the space and calculate Pearson correlation with all possible orthogonal combinations of the left-right with the other 32 suitable CHES issues of German political parties.\footnote{The CHES includes more issues, but we exclude those that express salience (that is, how important an issue is for a party).} 
In fact, we find that four CHES issues have a nearly equally high maximum correlation (all $>0.91$). These are: elite-skepticism (called `people vs. elite' in CHES), general position towards EU integration, position on EU's internal market and trade protectionism. We provide the detailed CHES description of these issues in the Appendix. The four CHES issues' corresponding optimal rotations are all within a window of only 22 degrees. Hence, we build an issue-combined second axis by averaging over their rotations of maximum correlation. Even with this averaged rotation, Pearson correlation with each of the issue dimensions is higher than 0.9 each. (Left-right has a pearson correlation of 0.94 with party positions on the x-axis, then.) We tested inclusion of arbitrarily many CHES issues with a cutoff of a Pearson correlation of 0.8. The next issue in line would have been `anti-islam rhetoric', but this issue dimension only achieved a correlation of 0.79. We use this cutoff because it implies strong positive relationship between the direction in the space and the issues. While such a fixed cutoff is still arbitrary, we aim to err on the safe side here: It allows us to safely assume that all four CHES issues are sufficiently well-represented by the second dimension, and its interpretation is not diluted by including as many CHES issues as possible.

We center the space around the mean user position along each dimensions (indicated by 0). To be precise, we also mirror the space at $y=0$. This is not strictly necessary, but if we do that, the more elite-skeptical/EU-skeptical/protectionist parties are placed further \textit{up} the y-axis, which facilitates interpretation of the plot. 

With this procedure, we arrive at a CHES-augmented two-dimensional political space. It provides one left-right axis and one axis representing degree of elite-skepticism, EU-skepticism, position towards EU's internal market and protectionism. In the following, we will sum up the latter four CHES dimensions under the term elite-/EU-skeptical protectionist (sometimes abbreviated as EESP).\footnote{The position towards EU's internal market is implicitly included in the combination of economic protectionism and general positions towards EU integration.} The further up the y-axis of the plot a party is positioned, the \textit{more} elite-/EU-skeptical/protectionist it tends to be.

The political positioning of users and parties in the space represents the basis on which a political cartography of news sharing rests. While not being a result in the strict sense, but rather the foundation on which news sharing can be studied, we report it with the results in the following section. Before, we turn to the description of data collection and content analysis.
%
%
%
%
%
%
\subsection{Link collection}

For a broad overview of which articles are shared by users placed in different regions of the political space, we composed a list of 26 German news outlets such as \texttt{faz.net}, \texttt{nachdenkseiten.de}, \texttt{bild.de} or \texttt{taz.de} (for the full list and inclusion criteria, see Appendix). In this list, we included both major legacy news outlets of different political leaning (so-called `Leitmedien' \citep{weischenberg2006journalismus}) as well as `alternative' media outlets \citep{schwaiger2022gegen}. 

We collected all tweets that contained a link to any one of those outlets for the month of March 2023. The tweet collection was carried out retrospectively with the Twitter Research API at the beginning of April 2023. The data set also includes shortened links (such as \texttt{bit.ly/AAAAA}) that do not contain the URL snippet of the respective outlet in the tweet, but do so once they are unwound again.
Our data set is made up of more than 500,000 unique tweets that contain a link to at least one of the news outlets. It both comprises direct shares (i.e., a user posted a tweet with a link to a news article) and retweets (i.e., a user retweeted a tweet by another user that contains a link).\footnote{In the pipeline at hand, we are interested in news circulation among users of differing political leaning. Hence, we do not distinguish between shares and reshares. It might be insightful to test whether news circulation of certain partisans is more or less reliant on few intermediaries.} 

Not all users who share a news article appear in our embedding. Yet, coverage is relatively high: Of all links that were shared by users that were not newspaper accounts, 60 percent came from users that follow at least three MPs.

\subsection{Article full texts, topic model, and metatopic assignment}

To assess news content, the texts of the articles beyond headlines and teasers were required. Using the links in the set of tweets, we automatically crawled the full text of the corresponding articles. This was done with the \texttt{newspaper3k} Python library \citep{n3k-ouyang}. We were able to collect roughly 60,000 unique article texts, corresponding to nearly 75 percent of all shared links. Since paywall policies differ strongly between outlets (with some not having any paywalled articles at all), the share of retrieved full texts was not equally distributed. We provide these statistics in the Appendix.\footnote{Outlet- and story-level analyses can be carried out regardless of how many articles have been retrieved, but one has to be careful once topics come into play, especially when intersecting topics with outlets. For in-depth topical analyses, it might be sensible to check the paywall policies of the respective newspapers; if certain topics are explicitly mentioned there or are implicitly overrepresented in paywalled articles, this might skew analyses. To assess the latter, one can test a randomly chosen subset of headlines of non-retrieved articles with respect to their topic distribution.}
We then followed a semi-automated approach. First, we inferred 220 topics based on a Structural Topic Model (STM) \citep{roberts2019stm}. Additionally, we categorized the topics into more coarse-grained metatopics. 

For topic modelling, we first removed stop-words and URLs. We lemmatized and tokenized the articles using \texttt{spaCy} \citep{honnibal2020spacy}, treating named entities and compound nouns as single tokens. This gave us a bag-of-words representation of each article. We then inferred a Structural Topic Model (STM) using the \texttt{stm} library in R \citep{roberts2019stm}, giving us 220 topics in the studied corpus (for a validation of the topic model, see Appendix). An article was considered to contain a given topic if the topic probability was higher than 0.5.\footnote{We choose this threshold because (i) it is sufficiently high to exclude too much noise from topics, (ii) because the distribution of maximum topic probabilites per article peaks at around 0.5 such that a reasonably high number of articles gets assigned a topic (for the distribution, see Appendix), and (iii) the maximum sum of all topics belonging to a metatopic per article enters a plateau at 0.5 (see Appendix, as well), such that most articles get assigned a metatopic.}

Subsequently, each of the 220 topics were classified into one of 12 metatopics (such as international politics, national politics, sports, or science/technology/health) by two of the authors according to a codebook (see Appendix).\footnote{A human in the loop allows oversight of the otherwise inductive, automated labelling procedure. Additionally, this ensures continuity with previous research done on news topics online, for which this has been carried out in this fashion (see \cite{quandt_no_2008}).} This was done by observing both the most relevant words for the topic (using a relevance score that computes the ratio between the log frequency of the word in the topic and the log frequency of the word in other topics), and by reading the top 3 articles that had the highest probability of containing that topic.\footnote{This allowed (i) to check whether the relevance score conformed with the most fitting articles for the topics and (ii) was manageable for human coders for all 220 topics.} The codebook for the metatopic classification was based on \cite{quandt_no_2008} and \cite{reuters-2023}. We use this number of metatopics because we aim for a coarse-grained classification of content here which is also supposed to be comparable internationally -- this was also the rationale in \cite{quandt_no_2008} on which be largely base our metatopics.

The basic idea behind our approach is the following: It is, quite obviously, unfeasible to classify such a large number of articles manually. The topic model allows automated topic inference into quite highly resolved topics. For example, news on Germany sending certain types of weapons to Ukraine are separated from general news on the Russian aggression against Ukraine. If a researcher is interested in these fine-grained content distinctions, the topic model enables analyses on this level. 
Yet, an investigation of whether users share only specific contents from an outlet is still unmanageable: It would require analysing thousands (220 times 26) of sharing distributions in our case. But it can be done if one uses 12 more coarse-grained metatopics. Score keywords for the individual topics as well as their metatopic assignment can be found in the Appendix.


\section{Results}

\subsection*{Baseline embedding}

The baseline embedding is given in Fig. \ref{fig:ca-ches} (rightmost plot). As already noted, we only provide party positions as reference points, not all MPs. This facilitates readablity. Parties are indicated by their typical party color. We observe a pronounced gap between \textit{AfD} (blue), the right-wing populist party with which a coalition has been ruled out by all other parties in parliament, and the rest of the parties. \textit{AfD} occupies the right-most and most elite-/EU-skeptical/protectionist position in the space. For the other parties, a more gradual transition can be observed. \textit{Die Linke}, the left-wing party of German federal politics (pink), is taking up the leftmost position. Then, \textit{Buendnis '90/Die Gruenen} (the Green party, green), \textit{SPD} (Social Democrats, red), \textit{CDU/CSU} (center-right Christian Democrats, black) and \textit{FDP} (market-liberal Free Democrats, yellow) are arranged. Within this group of parties, \textit{FDP} occupies the rightmost position, as well as the lowest value on the EESP axis (\textit{Die Linke} the highest).

Users, on the other hand, are represented as a green density plot. We observe that most users are embedded between \textit{SPD} and \textit{CDU/CSU}, that is, close to what is commonly understood as the political `center' of German politics. While user density gradually decreases around this center of density, another, yet much smaller peak is located close to \textit{AfD}.
We now showcase that and where the proposed research pipeline can substantially further our understanding of news sharing on digital platforms. 


\subsection{One-dimensional vs. two-dimensional political spaces}

\begin{sidewaysfigure}
    \includegraphics[width = \textwidth,keepaspectratio]{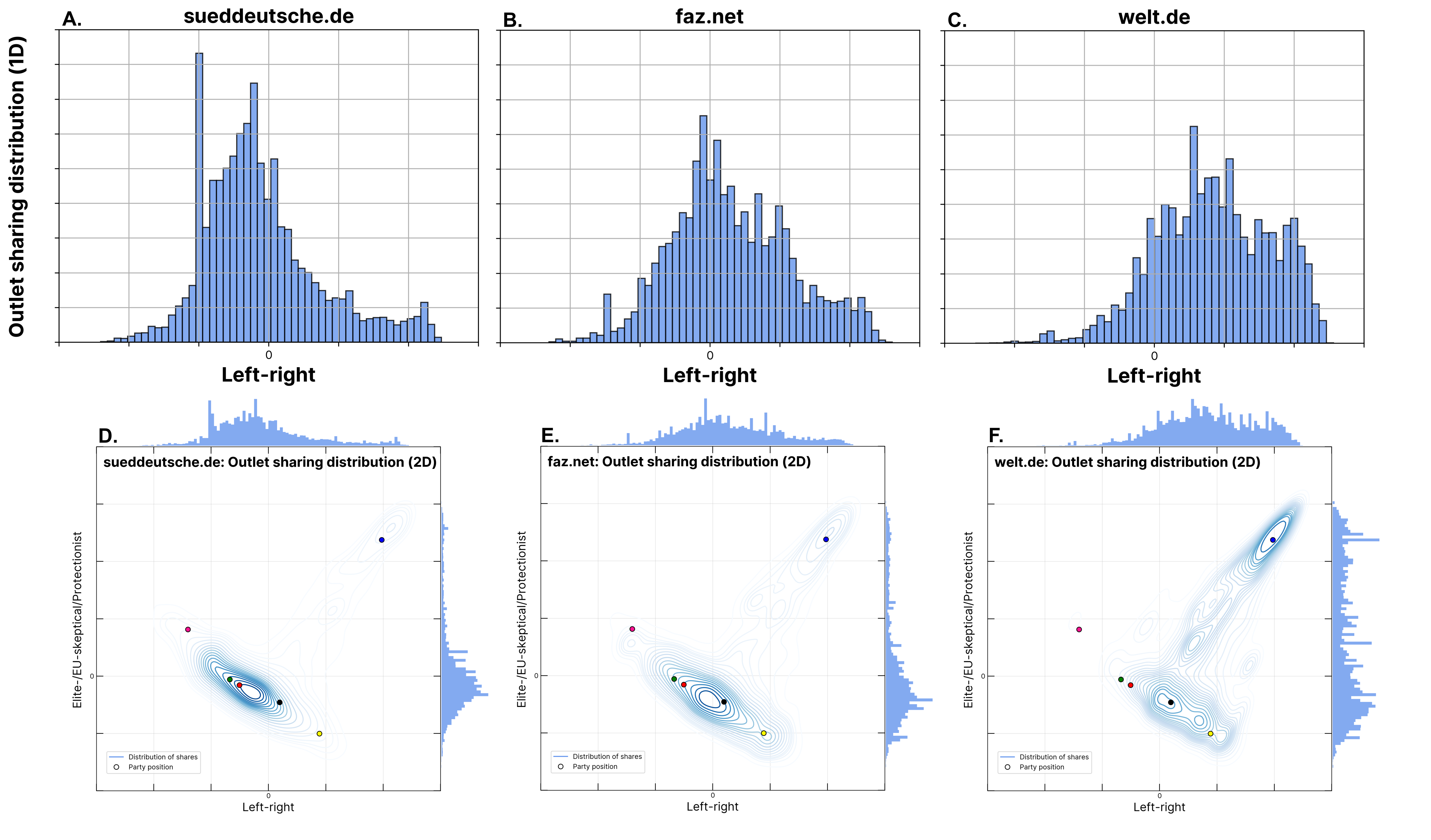}
    \caption{One-dimensional (A.-C., along the commonly-used left-right axis) and two-dimensional (D.-E.) outlet sharing distributions for three example outlets, namely \texttt{sueddeutsche.de}, \texttt{faz.net} and \texttt{welt.de}.
    Results exhibit fundamental differences. In the one-dimensional view, the differences between outlets appear gradual. The two-dimensional view reveals that \texttt{sueddeutsche.de} and \texttt{faz.net} are much more similar than \texttt{faz.net} and \texttt{welt.de}. The latter circulates in political regions at the other side of a fundamental political fault line along the EESP axis.
    }
    \label{fig:1dvs2d}
\end{sidewaysfigure}

The most relevant question concerning the two-dimensional political space is: Does it add relevant information, i.e., systematic differences in news sharing, that slip through the cracks in the one-dimensional view?

In fact, it does. We show in Fig. \ref{fig:1dvs2d} that differences between outlets that appear gradual along the commonly used one-dimensional left-right axis actually turn out to be more fundamental if an additional axis is considered. In the one-dimensional view (A.-C.), \texttt{sueddeutsche.de} is commonly shared left of center, while \texttt{faz.net} peaks around the mean of the left-right axis, but its distribution is skewed to the right. \texttt{welt.de} appears to be shared by users not too dissimilar from the ones sharing \texttt{faz.net}, yet the peak of its sharing distribution can be found right of center, and there are comparably more shares in the far-right region. 

The two-dimensional political space yields a starkly different impression (D.-E.).\footnote{Note that the distributions at the upper margins of the two-dimensional distributions are, by definition, equal to the ones shown in Fig. \ref{fig:1dvs2d} A.-C.. The minor differences that are visible result from different bin sizes which we used to increase readability of the one-dimensional distribution.} 
\texttt{welt.de} is the only outlet of the three that receives a significant amount of shares from the far-right, elite-/EU-skeptical/protectionist area of the space. The two-dimensional view reveals that \texttt{sueddeutsche.de} and \texttt{faz.net} have way more commonalities in terms of which user base distributes them than \texttt{faz.net} and \texttt{welt.de}. The latter circulates, to a considerable extent, in political regions at the opposite side of a fundamental political fault line along the EESP axis. H1 can be answered positively.

\subsection{Outlet-only analysis vs. including individual stories/content}

\subsubsection{Story level}

Simply investigating which outlet is circulated, e.g., on different sides of the political spectrum, does not mean it is \textit{the same content} that is shared by these different partisans. To address this question, one needs to focus on users sharing individual stories (as is also argued in \cite{green_curation_2021}).

\begin{figure}
    \includegraphics[width = 13.4cm,center]{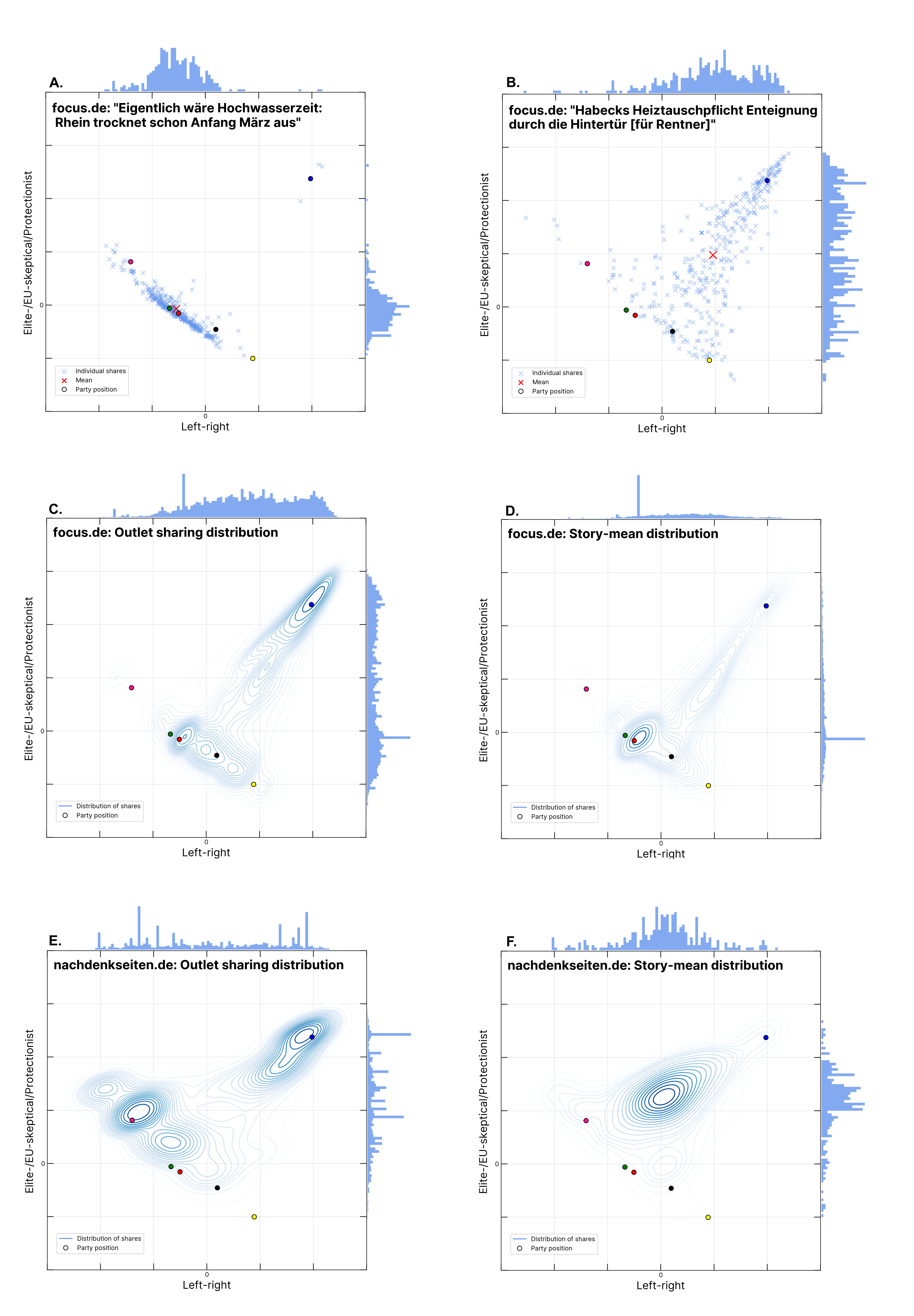}
    \caption{Circulation of individual news stories (A., B.), outlet-level distributions of two outlets (C., E.) and distribution of story means (D., F.). 
    Two articles from \texttt{focus.de} are shared in different regions of the political space (individual shares are displayed as blue crosses, the mean of all shares of the respective story by a red `X'). If we investigate the distribution of users sharing \texttt{focus.de} articles (regions of high density are darker, and since we deal with sharing and not with the baseline distribution now, we color the distribution blue), we see that the outlet is both shared by users placed close to \textit{AfD} as well as by center-right to right, less elite-/EU-skeptical/protectionist users (C.). Still, if the distributions of \textit{individual story means} is considered, we find that articles by the outlet rarely bridge the latter divide. They are either shared on one side \textit{or} the other of the cleavage (D.). An outlet that succeeds in publishing articles being shared on \textit{both} sides is \texttt{nachdenkseiten.de}: Most mean story positions are between \textit{AfD} and \textit{Die Linke} (F.).
    }
    \label{fig:story-outlet}
\end{figure}

In the framework we present here, it is possible to visualize the positions of the users that shared specific articles (see Fig. \ref{fig:story-outlet} A. and B.). An article by \texttt{focus.de} on the river Rhine running dry early in the year (A.) was shared mostly by users placed in the center or left of center in the less elite-/EU-skeptical/protectionist area of the political space. On the other hand, critical reporting on Germany's Economy and Climate Protection Minister Robert Habeck's law on building energy was shared mostly to the right of center -- and to the largest part in the right-wing and elite-/EU-skeptical/protectionist area of the embedding. 

The average position of the users sharing the respective article in Fig. \ref{fig:story-outlet} A. and B. is indicated by a red `X'. We can calculate this average for \textit{all} articles of a specific outlet, such as \texttt{focus.de}, and compare it to the distribution of users that share articles of the outlet. Graphically, the former amounts to a distribution of the average sharing positions (red crosses) of all \textit{Focus} articles, the latter to all blue crosses (of all the outlet's articles).
\texttt{focus.de} generally has a multi-modal sharing distribution (C.). 
If users across the elite-/EU-skeptical/protectionist cleavage would share the \textit{same} articles, then this would become visible as one large peak about halfway between the multiple peaks in C. -- their averages would be placed mostly between the `poles'. But this is not the case. For the most part, individual articles are either shared on one or the other side, as is visible in Fig. \ref{fig:story-outlet} D..\footnote{For \texttt{focus.de}, this heterogeneity can at least partly be explained by the fact that partisans differ in terms of which topics they share from the outlet. News on environment and climate are shared mostly by less EESP users (see Fig. \ref{fig:outletxtopic}), news on crime mostly by right-wing EESP users (see Appendix).} An outlet that indeed publishes articles shared both by left- and right-wing users is the alternative news page \texttt{nachdenkseiten.de}, where we see this large peak between the poles (F.). 

If only the outlet distribution (or mean position) was considered, such vital information would be overlooked: \texttt{focus.de} and \texttt{nachdenkseiten.de} appear to exhibit very similar sharing patterns in Figs. \ref{fig:story-outlet} C. and E. -- just that the former is shared across the EESP cleavage mostly by users right of center, whereas the latter by either left- or right-leaning users. But \texttt{nachdenkseiten.de} produced content that is shared by both sides of the EESP cleavage \textit{at the same time}, while for \texttt{focus.de}'s articles, it is mostly either-or. Whether an outlet produces content of interest to a broad set of users or just supplies different political camps with content in an `unbundled' way can only be assessed if the difference between story and outlet level is taken into account. H2a is confirmed. 

\subsubsection{Content level}
News circulation at the intersection of outlet and content level is highly interesting and has not been studied systematically with digital trace data. 

It is at this level where we can see unbundling at play. There are striking examples of outlets being shared in a region of the space if they cover one metatopic, and nearly completely ignored by the same users if they cover another one.\footnote{This analysis could also be carried out for outlet-topic combinations, too. The problem is that displaying all possible topic-outlet combinations would amount to 220 (number of topics) times 26 (number of outlets) plots and is hence barely feasible. Still, in principle, including outlet-topic combinations is possible.}

\begin{figure}
	\vspace{-1cm}
	\centering
	\includegraphics[width = 16cm, center]{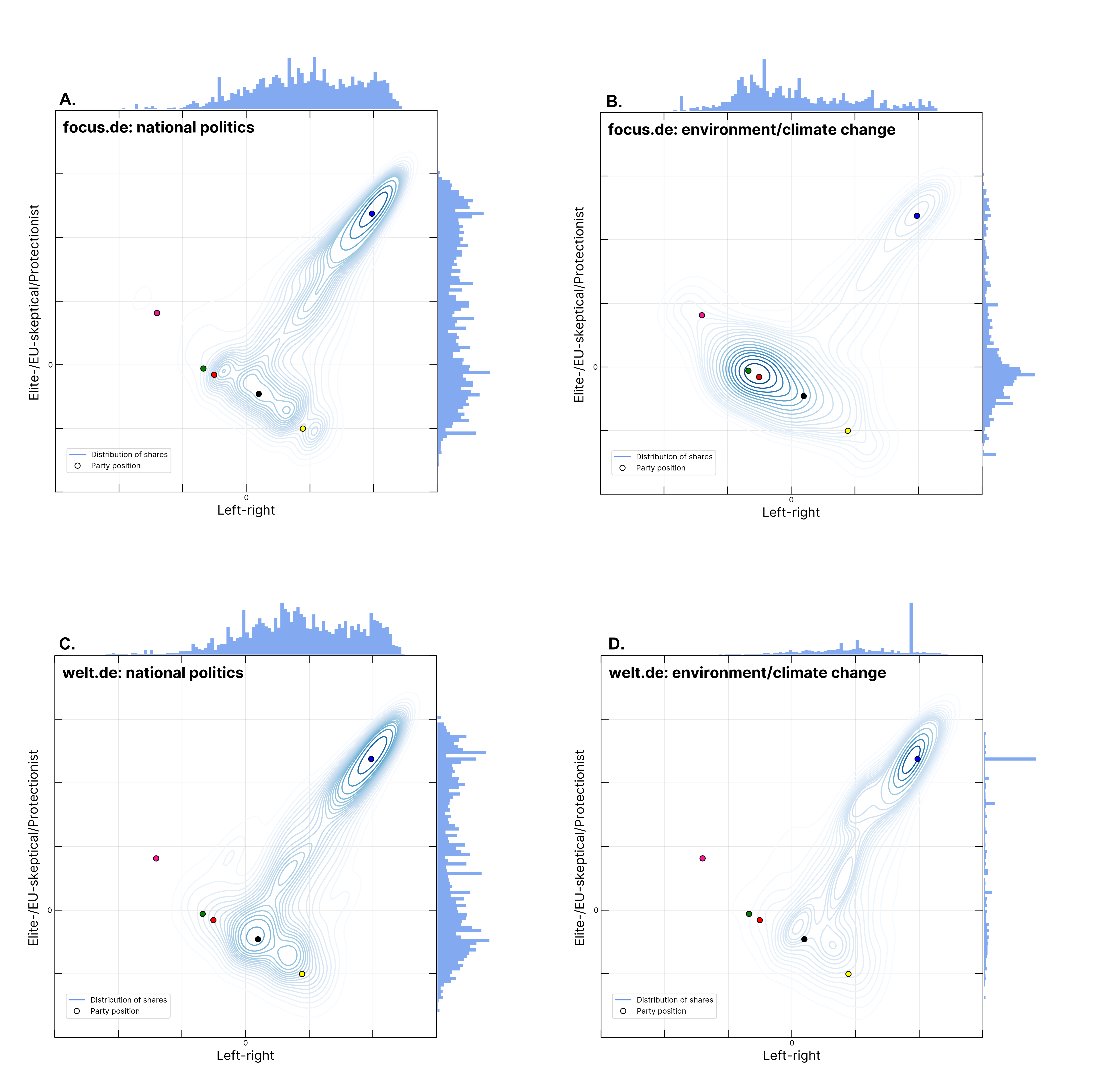}
	\caption{Comparison of the sharing distributions of \texttt{welt.de} and \texttt{focus.de} for different metatopics. 
		While articles on environmental issues and climate change by the latter outlet are shared comparatively more often by users left of center (B.), its coverage on national politics is shared by right-leaning users both high and low on the EESP axis (A.). \texttt{welt.de} articles on national politics (C.) are very similarly distributed to \texttt{focus.de}'s articles on the same topic. News on climate change and environmental issues, on the other hand, are shared to a disproportionate extent by right-wing EESP users (D.).
	}
	\label{fig:outletxtopic}
\end{figure}

For example, articles by \texttt{focus.de} on national politics are shared mostly to the right of center, both by center-right as well as (even to a larger extent) by right-wing elite-/EU-skeptical/protectionist users (Fig. \ref{fig:outletxtopic} A.). On the other hand, news by the same outlet on environmental issues and climate change are shared comparatively more often by users left of center (we see a strong peak between \textit{SPD} and \textit{Greens}, Fig. \ref{fig:outletxtopic} B.) and nearly not at all in the right-wing elite-/EU-skeptical/protectionist direction. In comparison, \texttt{welt.de} articles on climate change and environmental issues are shared to by far the largest extent by these users (Fig. \ref{fig:outletxtopic} D.), while news on national politics are shared mostly by center-right to right-wing users (C.).\footnote{As we noted above, analyses that include the content level (especially if one intersects content with the outlets). \texttt{focus.de} is among the outlets with a full text retrieval rate of more than 95 percent; for \texttt{welt.de}, 60 percent of article shares could be assigned a full text, which is slightly less than the general retrieval rate. We hence checked the outlet's paywall policy which does not discriminate between topics.} 

In these examples, it becomes visible how selectively news is shared on social media. \textit{Die Welt} and \textit{Focus} are on average shared at nearly identical positions in our space. But as we observe here, that does not imply that the same users indiscriminately share the same issues these outlets report on. 

This approach helps to unravel that while users might have a relatively balanced news diet in terms of outlets they share, they might simply share specific parts of outlets' reporting. This can amount to much less diversity in content coverage than is suggested on the outlet level. H2b is confirmed, too.

\subsection{Top-down vs. bottom-up approaches}

Attempts to capture news circulation online without either ignoring or drastically simplifying the topic dimension (e.g., by distinguishing only between political and non-political news in a top-down approach) remain rare.\footnote{A notable exception is \cite{jurgens2022mapping}, but they deal with the effect of algorithmic curation on news diversity.} Our topic model yields 220 topics in the case at hand, which, content-wise, preserves considerable detail. Coarser metatopics are also provided. We now investigate the interplay of these two layers:
To this end, we observe the (hand-coded) metatopic of climate change and environment. It subsumes several topics identified by the topic model, of which we display a selection here: One on the science and institutions concerned with climate change, one on weather, especially extreme weather events, and two on climate protests: Fridays for Future, and the more extreme Last Generation (`Letzte Generation').

\begin{figure}
\vspace{-2cm}
    \centering
    \includegraphics[width = 14cm, center]{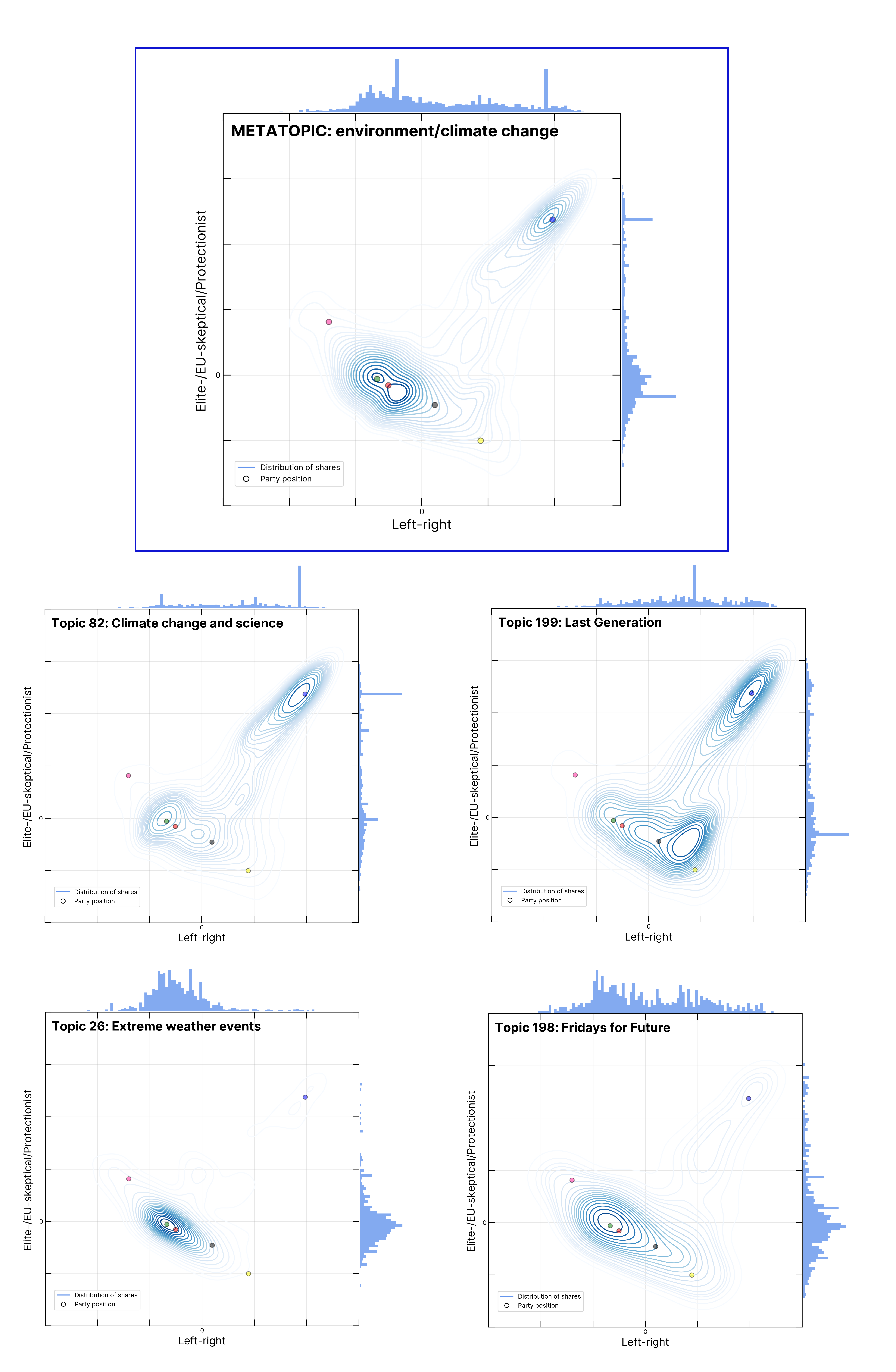}
    \caption{Four topics and their corresponding metatopic (environment and climate change). 
    News on environment and climate change are shared by users across the political spectrum. Yet, stories on extreme weather events and Fridays For Future are shared nearly only left of center. News on the more extreme climate movement Last Generation and on the science and institutions dealing with climate change are shared by the political counterparts (right of center and by right-wing elite-/EU-skeptical/protectionist users, respectively).
    }
    \label{fig:topic-metatopic}
\end{figure}

In Fig. \ref{fig:topic-metatopic} C., we see that news on environment and climate change are shared by users across the political spectrum. But by increasing `topic resolution', that is, taking a look at the topics that compose this metatopic, we find that the different sides in our political space are concerned with different aspects of it. News stories concerned with extreme weather events are very rarely shared in the right-wing elite-/EU-skeptical/protectionist area, while the distribution peaks left of center. On the other hand, articles on climate change more generally and the institutions concerned with it are circulated heavily in the former area, and a lot less so at the other end of the elite-skeptical dimension. News on the climate movement Last Generation are shared right of center across the EESP axis.

Hence, there is substantial heterogeneity in sharing patterns on the topic level. Sometimes, certain topics might not be of interest to specific user groups at all. In other cases, the topic-metatopic distinction can provide deeper insights into large-scale sharing patterns: as showcased above, only specific aspects of broader issues might be circulated in different political circles. Our approach can uncover this heterogeneity by classifying content at different levels of granularity, which can contribute to an understanding of how news stories polarize political discourse. H3 can be confirmed.

\section{Discussion}

We have argued that studies of news sharing often lack a broader, systematic view on the phenomenon. We have then pointed out three methodological concerns that, if addressed, allow a step towards such an overview. Subsequently, we have introduced an inductive, multi-scale approach to the study of news sharing on Twitter on the story, outlet and topic level. We sketched a research pipeline, which rests on three pillars: (i) An estimation of users' political positions in a CHES-validated, two-dimensional political space, based on follower network embeddings; (ii) an extensive collection of shares of newspaper articles; (iii) automated collection of said articles, along with (semi-)automated classification of their content. 

The resulting embedding exhibits, in the case of German MPs' Twitter follower networks, a triangular structure: MPs are ordered along their parties' positions on a left-right axis, but additionally, \textit{AfD} politicians are positioned, along with the users following them, quite far away from the other parties in the second dimension. This dimension represents degree of elite-/EU-skepticism, positions towards EU's internal market and protectionsim. The cleavage would have been overlooked if only a one-dimensional left-right axis had been considered. Knowing about this additional dimensions is important, however, because it explains a major political fault line of German politics.

Based on one month of sharing data of the German Twittersphere, we went on to showcase examples of smaller- and large-scale sharing patterns which become visible with this framework. They might be overlooked in approaches that are mostly outlet-based, top-down or employing a one-dimensional estimation of users' political leanings.

First, we present striking examples of sharing distributions that look similar in a one-dimensional political space but exhibit \textit{categorical} differences in the two-dimensional view. 
Secondly, we show that outlet level-only approaches overlook heterogeneities with respect to the circulation of individual stories. Additionally, combining content and outlet level, we observe that depending on the metatopic, certain outlets are preferred or ignored by specific user groups. For example, while right-leaning \texttt{focus.de} was generally shared often by right to right-wing users, its coverage of environmental issues was largely ignored especially by right-wing elite-/EU-skeptical/protectionist users. 
Thirdly, there is also substantial heterogeneity in sharing patterns on the content level. Sometimes, certain topics might not be of interest to specific user groups at all. But even if certain topics are shared by users across the political space, they might not focus on the same \textit{aspects}. The distinction between topic and metatopic can provide deeper insights into large-scale sharing patterns: Exemplarily, we found that news on climate change and environment were shared across partisan divides -- but right-wing elite-/EU-skeptical/protectionist users focused on reporting about institutions and research dealing with climate change, while users left of center often shared news on extreme weather events.

We hope that this research pipeline can impact and be adopted in future studies of online information circulation. Also partial extensions of standard research procedures -- may it be using multidimensional estimates of political leaning, taking into account the content level or using a broad spectrum of outlets and a bottom-up approach to classify topics -- can yield important additional insights. There are several ways in which studies that cannot adopt the full framework might use parts of the pipeline presented here.
If content is unavailable, the circulation of individual stories (which can be discerned by links and headlines) might already suffice to show whether news sharing is systematically unbundled or not. If users' political positions are estimated via surveys instead of interaction networks, additional questions (for example, on attitudes towards the EU and protectionist politics) can complement self-placement on a left-right axis. If consumption data is available instead of sharing data, studies can be carried out analogously to politically map news \textit{exposure}.

Of course, this method comes with limitations: Firstly, it is primarily tailored towards Twitter/X data. It has become increasingly difficult for researchers to get API access from X. It remains to be seen whether legislation, such as the EU's Digital Services Act, can render research with digital traces from X possible again.
For other digital platforms, it might either be more difficult to estimate user positions (e.g. for YouTube, Instagram), or to collect sharing data (TikTok, YouTube). Still, the method is applicable to these platforms in principle. The method of data collection will have to change. It might, for example, be advisable to use data donations  \citep{ohme2024digital} instead of an API. In the case of YouTube and TikTok, even news consumption could be studied since data donations include exposure data.
We carry out this study only for a single country, namely Germany. Due to the interplay between follower networks, CHES data and automated text processing, it can be used in a completely analogous way for all countries for which CHES data is available, and in principle, also more than two dimensions could be used. Germany presents a good example to test a two-dimensional space because right-wing populism is a relatively new phenomenon compared to other European countries (e.g. France). We find that the two-dimensional space indeed makes visible news sharing diversification along the second dimension.
Moreover, due to the embedding method relying on users following at least three MPs, roughly 60 per cent of all shares could be embedded in the political space. Therefore, we were not able localize all sharing events politically. Alternative ways of constructing a political space should be sought that might complement the approach. For example, on TikTok (or even X), likes could be used as signals of revealed preference if a user does not follow enough political actors.

This study makes clear that a systematic understanding of news sharing on digital platforms requires a broad set of outlets, a combination of multiple data sources, and both fine- with coarse grained levels of analysis. On this basis, we can make visible to what extent information circulation overlaps and differs across the political spectrum. It is the latter -- news fragmentation, informational asymmetries, problematic and conflictual news spreading -- upon which the focus of research on news sharing hitherto has mostly resided. But as, for example, \cite{mau2023triggerpunkte} stress in recent work, agreement is often surprisingly widespread with respect to several contested issues of German politics, yet stark conflict can nevertheless arise once certain groups see basic societal necessities threatened. The laws of the attention economy are strongly focused on such conflictual, divisive events. Yet, this should not blind us researchers ---a broader perspective is needed \citep{guess_consequences_2021,altay2023misinformation}. With respect to studies of news sharing, this extends to methodological breadth, as we have pointed out in this work. It does not matter if one studies moderates, partisans, or generally politically interested users: The news that are circulated cannot be measured based on one feature alone -- may it be that the news are problematic, or which outlet they come from, or which topic they cover. To arrive at well-formed hypotheses about news sharing (and consumption), one must consider the whole information environment on a digital platform. We attempted to sketch out an approach that can provide an advance in this direction.

\subsection*{Funding Statement}
FG and JO received funding from the `Digital News Dynamics' research group at Weizenbaum Institute for the Networked Society in Berlin, funded by the Federal Ministry of Education and Research. AP has received funding from the European Union's Horizon Europe framework (HORIZON-CL2-2022-DEMOCRACY-01-07) under project number 101094752.
\subsection*{Ethical approval and informed consent statements}
The public tweets collected in this study were analyzed and described in this paper following the GDPR. No personal information related to on any accounts were mentioned in the paper. The data were, as is most often the case in big data experiments, collected without the formal consent from the users that contributed to it. Even though their Twitter activity is public, they may not formally agree to be part of the scientific analysis \citep{rogers2018social}. Especially considering the sensitive nature of the presented computations, like the construction of a political space, which could be misused by nefarious actors for political profiling, we choose to systematically refrain from mentioning any kind of personal information that may lead to the identity of users. 
\subsection*{Data availability statement}
Twitter/X's Terms of Service do not allow us to share the full tweet objects of the collected tweets. We shall therefore share with other researchers the `hydratable' Twitter IDs upon request, which allows to retrieve the full tweet objects of tweets that still exist at the time of hydration. The code for the Correspondence Analysis and the rotation of the space is provided under \texttt{\url{https://github.com/fgais/political-sharing-cartography}}. 



\bibliographystyle{apalike}
\begin{spacing}{1.0}
\bibliography{bib}
\end{spacing}
\newpage
\appendix

\section{Outlet list}
Both major legacy and alternative outlets were included. The latter outlets are generally characterized by a self-understanding as a corrective vis-à-vis perceived hegemonic politics or media \citep[p. 174]{schwaiger2022gegen}. They more or less explicitly claim to publish and promote content that is unduly underrepresented in the mainstream public sphere and include ``both special-interest outlets that respond to a perceived lack of news diversity and outlets in an explicit opposition to a so-called `mainstream' within the media and political landscape'' \citep{strippel-2024}. We included major alternative news outlets in terms of reach and of differing political leaning, hence offering a broad understanding of alternative media as a corrective to a biased mainstream media system \citep{strippel-2024}.

\begin{tabular}{lll}
\toprule
Outlet & URL & Twitter handle \\
\midrule
BILD & bild.de & @BILD \\
Sueddeutsche Zeitung & sueddeutsche.de & @SZ \\
Frankfurter Allgemeine Zeitung & faz.net & @faznet \\
Handelsblatt & handelsblatt.com & @handelsblatt \\
Die Welt & welt.de & @welt \\
Focus & focus.de & @focusonline \\
Zeit (Online) & zeit.de & @zeitonline \\
Frankfurter Rundschau & fr.de & @fr \\
Stern & stern.de & @sternde \\
TAZ & taz.de & @tazgezwitscher \\
Spiegel & spiegel.de & @derspiegel \\
Epoch Times & epochtimes.de & @EpochTimesDE \\
PI-NEWS & pi-news.net & @P\_I \\
Journalistenwatch & journalistenwatch.com & @jouwatch \\
Die Achse des Guten & achgut.com & @Achgut\_com \\
Tichys Einblick & tichyseinblick.de & @TichysEinblick \\
Junge Freiheit & jungefreiheit.de & @Junge\_Freiheit \\
Deutsche Wirtschafts Nachrichten & deutsche-wirtschafts-nachrichten.de & @DWN\_de \\
NachDenkSeiten & nachdenkseiten.de & @NachDenkSeiten \\
Cicero & cicero.de & @cicero\_online \\
Neues Deutschland & nd-aktuell.de & @ndaktuell \\
Der Freitag & freitag.de & @derfreitag \\
junge Welt & jungewelt.de & @jungewelt \\
netzpolitik & netzpolitik.org & @netzpolitik\_org \\
Correctiv & correctiv.org & @correctiv\_org \\
Uebermedien & uebermedien.de & @uebermedien \\
\bottomrule
\end{tabular}

\section{List of MPs}
The basis for the collection was a list of 629 MPs in parliament for the federal legislature starting 2021 with registered Twitter accounts provided by the \textit{Epinetz} data set \citep{konig2022epinetz}. The follower network was then collected for all MPs that were actively participating in public debate in the German Twittersphere. To this end, we included all MPs that showed up in the retweet networks of at least 5 of around 2000 Twitter trending topics, collected daily from 2021 to 2023. This yielded 433 MPs. The follower network was collected in July 2023. For a list of these MPs, see separate \texttt{.csv} file.

\section{Correspondence Analysis}
\begin{figure}[ht]
\includegraphics[width=.8\textwidth]{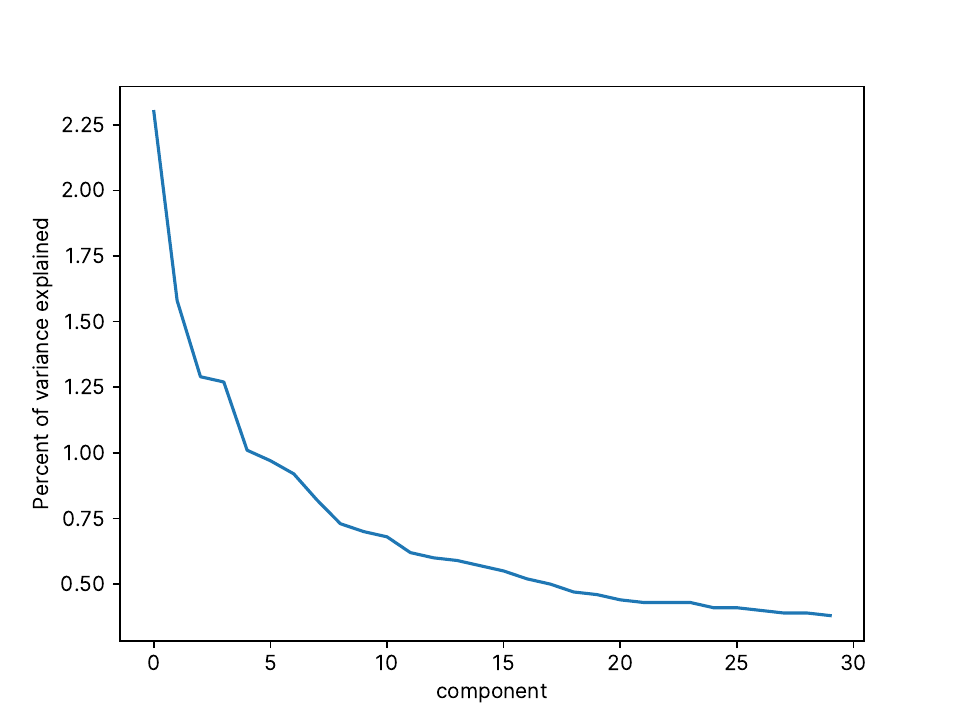}
\caption{\label{fig:ca_var}CA percent of variance per dimension.}
\end{figure}
The follower network can be described as a adjacency matrix $A$ with $N_{MP}$ columns and $N_{users}$ rows. An entry $a_{ij}$ is 1 if a user follows the respective MP, and 0 otherwise. The correspondence matrix $P$ is calculated by $P=\frac{A}{\sum_{i,j}a_{ij}}$. The matrix of standardized residuals $S$ is calculated via $$S=D_r^{1/2}(P-rc^T)D_c^{1/2},$$
with $r$ and $c$ being row and column masses ($r_i = c_j = \sum_j p_{ij}$, $D_r = \text{diag}(r)$, $D_c = \text{diag}(c)$) (see \cite{barbera2015tweeting}, Sup. Mat.). Then, Singular Value Decomposition is carried out such that $S = UD_\alpha V^T$ where $U^TU = V^TV = \mathbb{I}$. All rows and columns are projected onto that plane by then computing the standard coordinates $D_c^{1/2} V$ for MPs, $D_r^{1/2} U$ for other users \citep{barbera2015tweeting}.

As in \cite{ramaciotti_morales_inferring_2022}, we only included users that followed at least three MPs. But we take the full user-MP follower matrix, and not a reduced full-rank matrix. The reason for this is that a latent space inferred by only including unique connection patterns can turn out to be different from the original one, which becomes intuitively clear in a force-based formulation of latent space models: More users with the same following pattern increase the attracting force on the MPs they follow, and the repulsion on the others (see \cite{gaisbauer2023grounding}).

The code for the CA and the rotation can be found under \texttt{github.com/fgais/political-sharing-cartography}.

Percent of variance explained is displayed in Fig. \ref{fig:ca_var}.

\begin{figure}[ht]
\includegraphics[width=.8\textwidth]{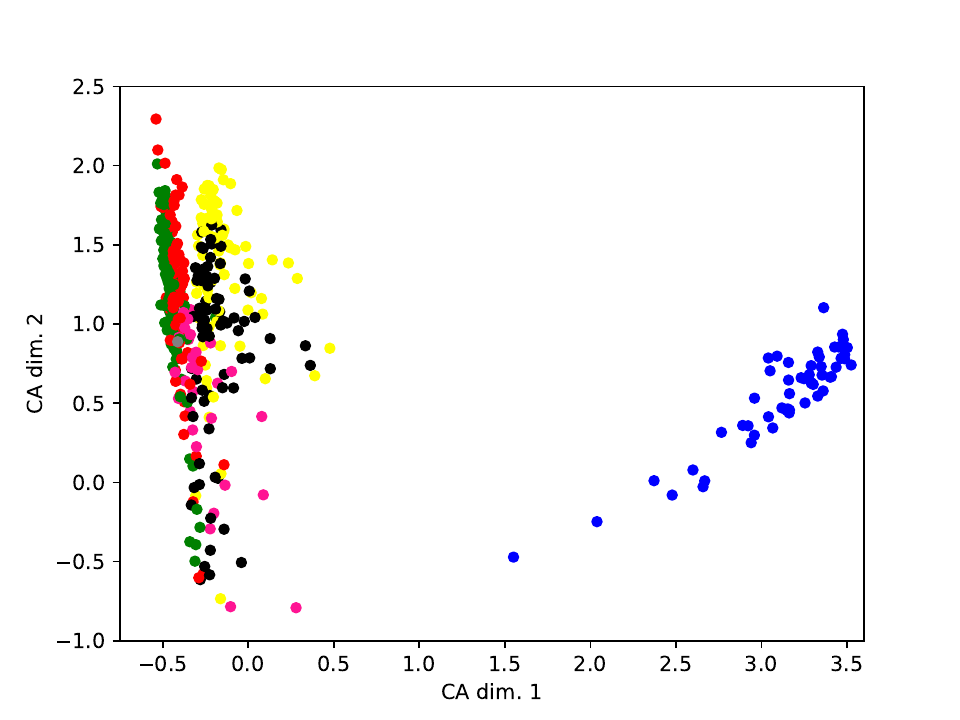}
\caption{\label{fig:ca_dim2}CA first and second dimension: MP positions for each party are disproportionately spread out along the second dimension, and the most important MPs (e.g., the party chairs or the ones with the most followers) tend to be placed close together, no matter the party.}
\end{figure}

Dimensions 1 and 2 of the CA are displayed in Fig. \ref{fig:ca_dim2}. As is visible, MP positions for each party are disproportionately spread out. In fact, the MPs placed far down on dimension 2 tend to be important party members of each party. We calculate Pearson correlation as well as Spearman Rank correlation with the MP's follower counts. We find relatively high correlation with follower counts, Spearman Rank 0.79, Pearson 0.58. None of the other axes shows such a high correlation (Pearson below 0.04 for the first and third Principal Component, Spearman Rank below 0.16 for each).

\section{CHES codebook entries}

\begin{itemize}

\item People vs. elite: ``[P]osition on people vs elected representatives. Some political parties take the position that `the people' should have the final say on the most important issues, for example, by voting directly in referendums. At the opposite pole are political parties that believe that elected representatives should make the most important political decisions.''

\item EU position: ``[O]verall orientation of the party leadership towards European integration in
2019''

\item EU internal market: ``[P]osition of the party leadership in 2019 on the internal market (i.e. free movement of goods, services, capital and labor).''

\item Protectionism: ``[P]osition towards trade liberalization/protectionism.''

All from \cite{ches_codebook2020}.

\end{itemize}

\section{Unretrieved articles}

\begin{figure}[htbp]
	\includegraphics[width=.8\textwidth]{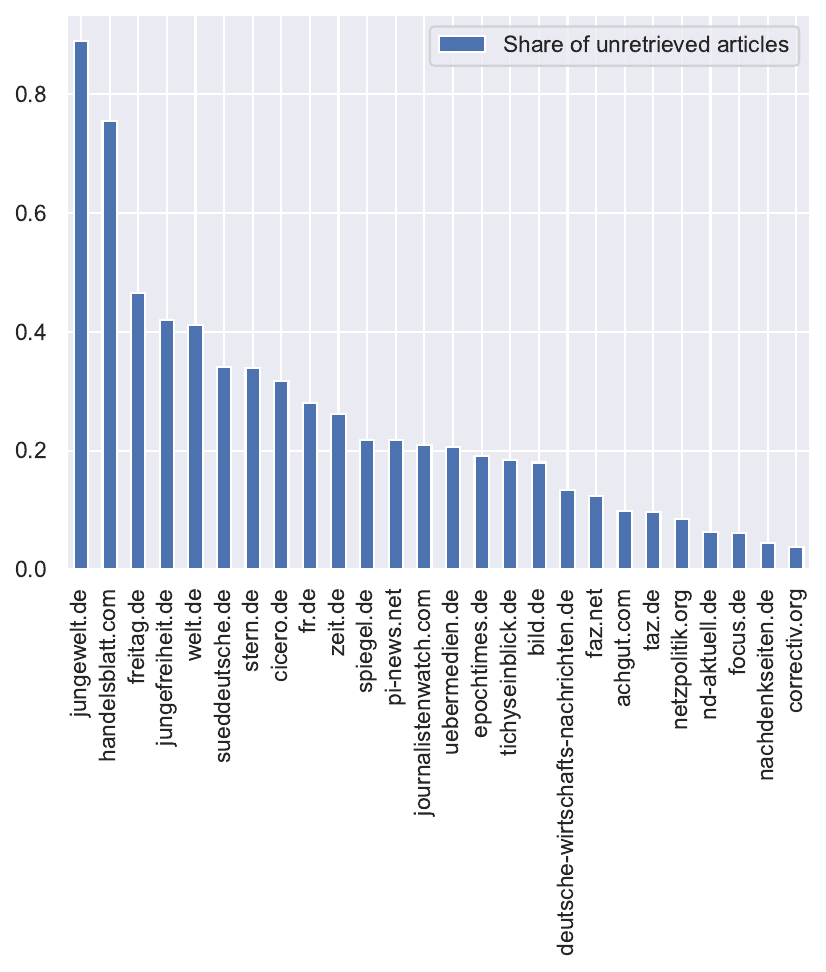}
	\caption{\label{fig:unretrieved} Share of unretrieved articles per outlet.}
\end{figure}

See Fig. \ref{fig:unretrieved}.

\section{Topic model selection}

\begin{figure}[htbp]
\includegraphics[width=.8\textwidth]{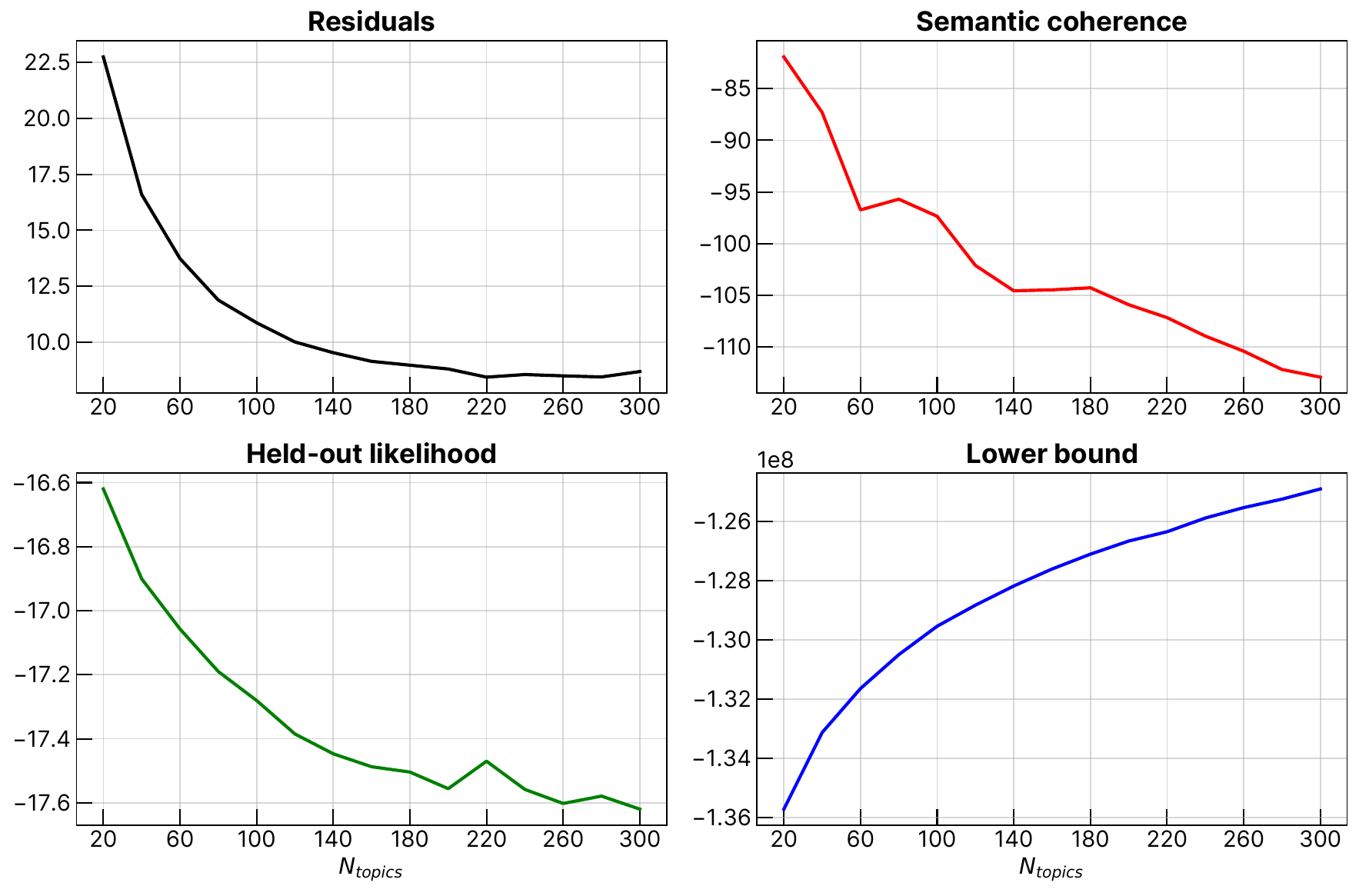}
\caption{\label{fig:stm_eval}Topic model evaluation scores.}
\end{figure}

The preprocessing of the raw articles consists of stop-word and URL removal. Then, the articles are tokenized and lemmatized using the Python library \texttt{spaCy}, treating named entities and compound nouns as single tokens. The sentence ``John Doe ran out of dishwashing soap'' thus would turn into ``[john\_doe,run,dishwashing\_soap]''. After preprocessing, we infer a structural topic model using the R library \texttt{stm}. The important hyperparameter here is the number of topics $N_{topics}$, which we vary between 20 and 300 in steps of 20, as can be seen in Fig.\ref{fig:stm_eval}. We then evaluate the different models according to standard model evaluation scores: residuals, semantic coherence, held-out likelihood and lower bound. None of these scores should be considered alone to select $N_{topics}$. Instead, they point the researcher to sensible parameter regions to examine manually. Considering the inflection points in semantic coherence and held-out-likelihood in Fig.\ref{fig:stm_eval}, we manually examined the models for 80, 100 and 220 topics. The latter yielded the best tradeoff between semantically coherent and fine-grained topics, which led us to choose it for the further analysis.

\section{Codebook for metatopics}

The metatopics themselves are based on \citep{quandt_no_2008} and \citep{reuters-2023}. Coding was done independently for each topic by two coders, who, in case they assigned a topic different metatopics, settled on a common metatopic upon discussion. Initial Cohen's Kappa was .69. The topics along with lists of most representative words, as well as a list of their assigned metatopics, can be found in the separate text files provided.

\begin{itemize}
\item Environment/Climate Change:
\begin{itemize} 
\item Score keywords refer to (extreme) weather events, nature, natural disasters, or climate change
\item 3 most representative articles mention said terms and information related to them, in particular with a focus on the natural implications (as opposed to social implications)
\end{itemize}
\item Culture/Arts/Media: 
\begin{itemize}
\item Score keywords refer to cultural events or artists, works of art and/or cultural entertainment such as films, series or books
\item 3 most representative articles mention said terms and information with a focus on their cultural significance (not on celebrity news/gossip)
\end{itemize}
\item Sports: 
\begin{itemize}
\item Score keywords refer to sports-related actors/events
\item 3 most representative articles mention sports-related information with a focus on their significance for their sports disciplines (not on celebrity news/gossip)
\end{itemize}
\item Lifestyle/Human Interest/Celebrities:
\begin{itemize} 
\item Score keywords refer to stars, cultural and public figures, or lifestyle with entertainment connotation
\item 3 most representative articles provide information that is primarily entertaining or privately useful, e.g. the private life of stars, cultural and public figures, or DIY content
\end{itemize}
\item Crimes/Personal Security: 
\begin{itemize}
\item Score keywords refer to criminal events, courtroom converage, or personal security-related issues/events (e.g. accidents)
\item 3 most representative articles mention criminal events and investigations, courtroom coverage, or personal security-related issues
\end{itemize}
\item Economy/Finance/Business: 
\begin{itemize}
\item Score keywords refer to businesses, economic actors (that are not directly tied to political institutions/decisions), financial markets
\item 3 most representative articles mention economic or financial actors or developments (that are not directly focussing on political institutions/decisions)
\end{itemize}
\item Local: 
\begin{itemize}
\item Score keywords refer to actors, events or issues of local relevance
\item 3 most representative articles mention issues of only local relevance, such as news on city planning or regional events
\end{itemize}
\item Science/Technology/Health: 
\begin{itemize}
\item Score keywords refer to technology, science, or health research/news
\item 3 most representative articles focus on scientific or technological developments, science journalism, or health-related news
\end{itemize}
\item National Politics: 
\begin{itemize}
\item Score keywords refer to political or state institutions, politicians, or political events such as elections
\item 3 most representative articles focus on national political issues which are directly tied to political/state institutions, politicians, or political events (such as elections)
\end{itemize}
\item International/Foreign Politics: 
\begin{itemize}
\item Score keywords refer to international political issues, institutions, but also national political actors or events of other countries than Germany
\item 3 most representative articles focus on international political issues, especially conflicts (such as the war in Ukraine), institutions (such as the UN), but also national politics of other countries than Germany
\end{itemize}
\item Social Affairs: 
\begin{itemize}
\item Score keywords refer to societal issues, e.g. topics of education, migration, or employment and poverty (can be political, but not directly tied to political/state institutions)
\item 3 most representative articles focus on societal issues, e.g. topics of education, migration, or employment and poverty (can be political, but not directly tied to political/state institutions)
\end{itemize}
\item History:
\begin{itemize}
\item Score keywords refer to historical events
\item 3 most representative articles focus on a portrayal of historical events in a both informative and entertaining fashion
\end{itemize}
\end{itemize}

\section{Topic model probability threshold}
\begin{figure}[t]
	\includegraphics[width=.9\textwidth]{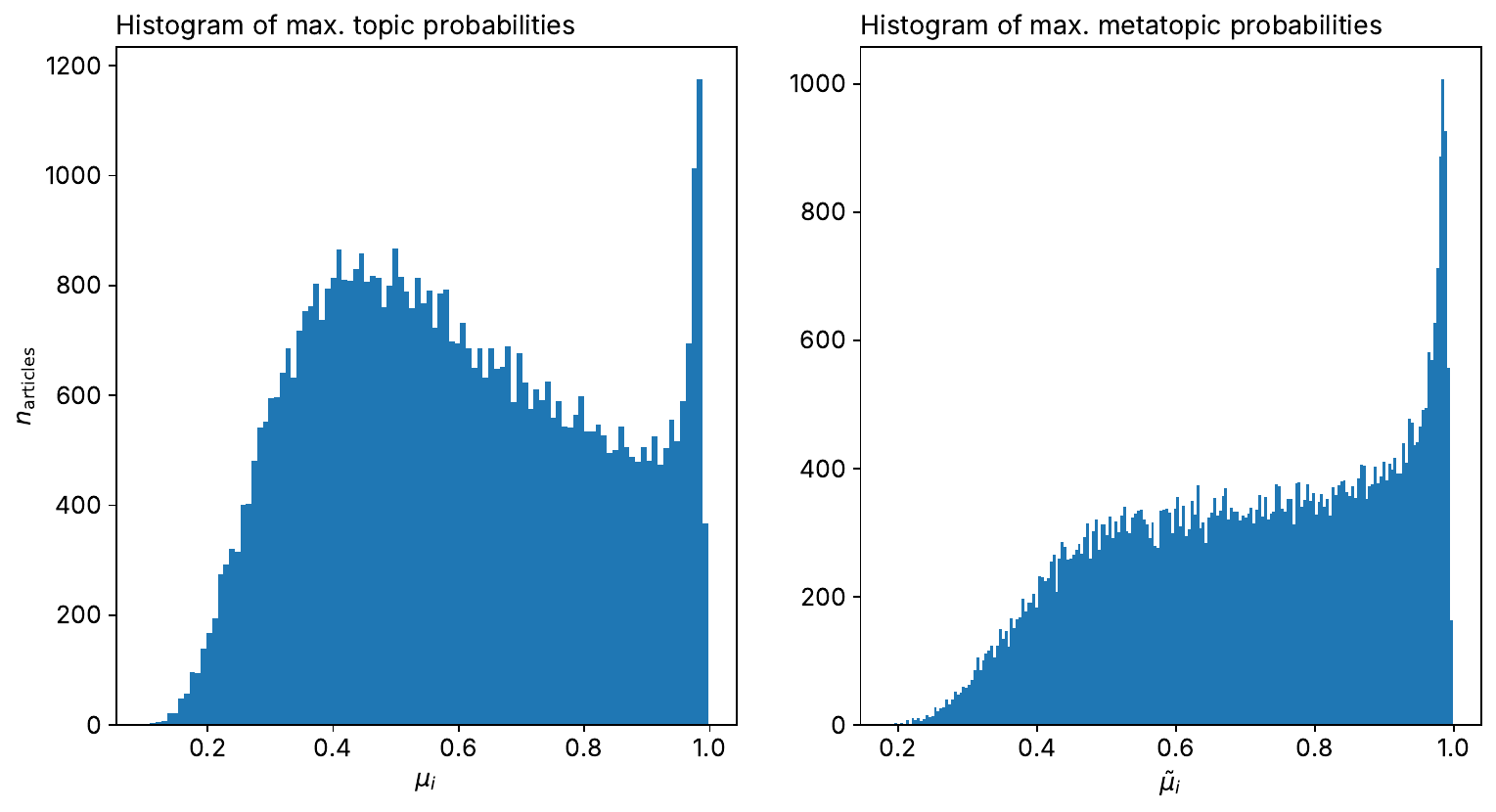}
	\caption{Histograms of max. topic probabilities $\mu_i$ (left) and max. metatopic probabilities $\tilde{\mu}_i$ (right).}
	\label{fig:topicmaxhist}
\end{figure}
Let $\mathcal{T} \in \{1,...,n_{\mathrm{topics}}\}$ be the topic index set. Let $\mathbf{M} = (M_{ij})$ be the document-topic matrix such that $M_{ij}$ is equal to the probability of topic $j$ in document $i$.
After grouping topics into metatopics as described in Sec. 3.3, we define the document-metatopic matrix $\tilde{\mathbf{M}} = (\tilde{M}_{ij})$ such that $\tilde{M}_{ij}$ is equal to the probability of metatopic $j$ in document $i$. This probability is given by
$$
	\tilde{M}_{ij} = \sum_{j' \in \mathcal{T}_j}M_{ij'}
$$
where $\mathcal{T}_j \subset \mathcal{T}$ is the set of topics that belong to the metatopic $j$.\\
In order to examine the distribution of (meta-)topics across the political space in Sec. 3.3, we define a main topic to each document. For each document $i$, we compute the maximum of its topic probability distribution $M_{ij}$ denoted as $\mu_i$:
\begin{equation}
	\label{eq:2}
	\mu_i = \max_i(M_{ij})
\end{equation}
Equivalently, we define $\tilde{\mu}_i$ as the maximum of the metatopic probability distribution of topic $i$. Fig.~\ref{fig:topicmaxhist} shows the histogram of the values of $\mu_i$ and $\tilde{\mu}_i$. For topics (left plot), we observe that the histogram peaks at a value around 0.5. For metatopics (right plot), we observe that the histogram reaches a plateau around a value of 0.5. These observations, combined with a close reading of selected articles with different values of $\mu_i$, leads us to formulate the following rule for assigning the main topic to a document: If $\mu_i > 0.5$, we assign the corresponding topic as the main topic to document $i$. Else, we disregard the document for this part of the analysis. Equivalently, we apply the rule $\tilde{\mu}_i > 0.5$ to assign the main metatopic. Following this scheme, we can assign a main topic to 62\% of the documents and a main metatopic to 81\% of the documents in the corpus. The threshold therefore provides a reasonable tradeoff between accuracy and sample size.

\section{Crime/Personal Security news from focus.de}
\begin{figure}[htbp]
	\includegraphics[width=\textwidth]{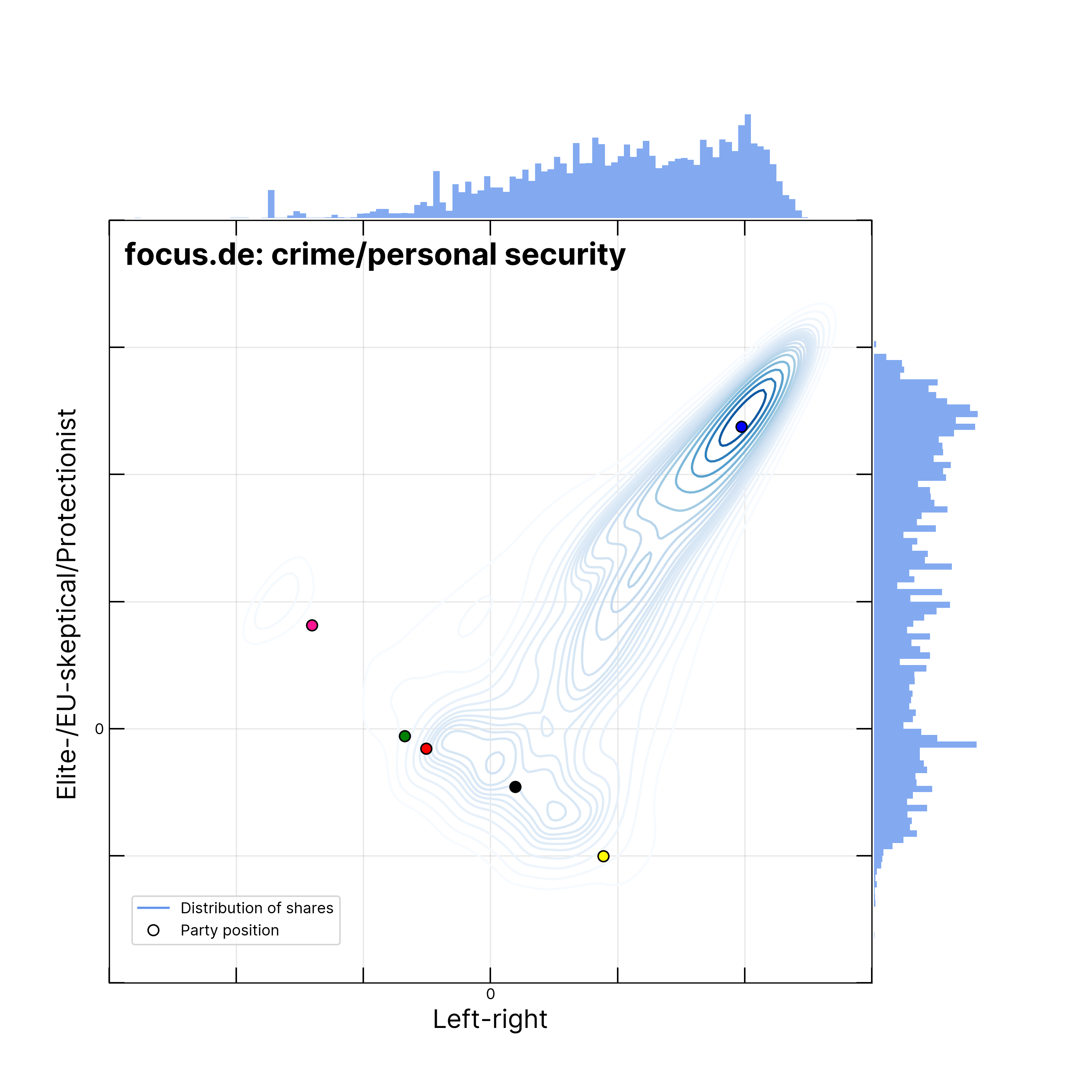}
	\caption{\label{fig:focuscrime}Distribution of news by \texttt{focus.de} on Crime/Personal Security.}
\end{figure}

See Fig. \ref{fig:focuscrime}.

\section{Other topics attributed to metatopic `Environment and Climate Change'}
\begin{figure}[htbp]
\includegraphics[width=\textwidth]{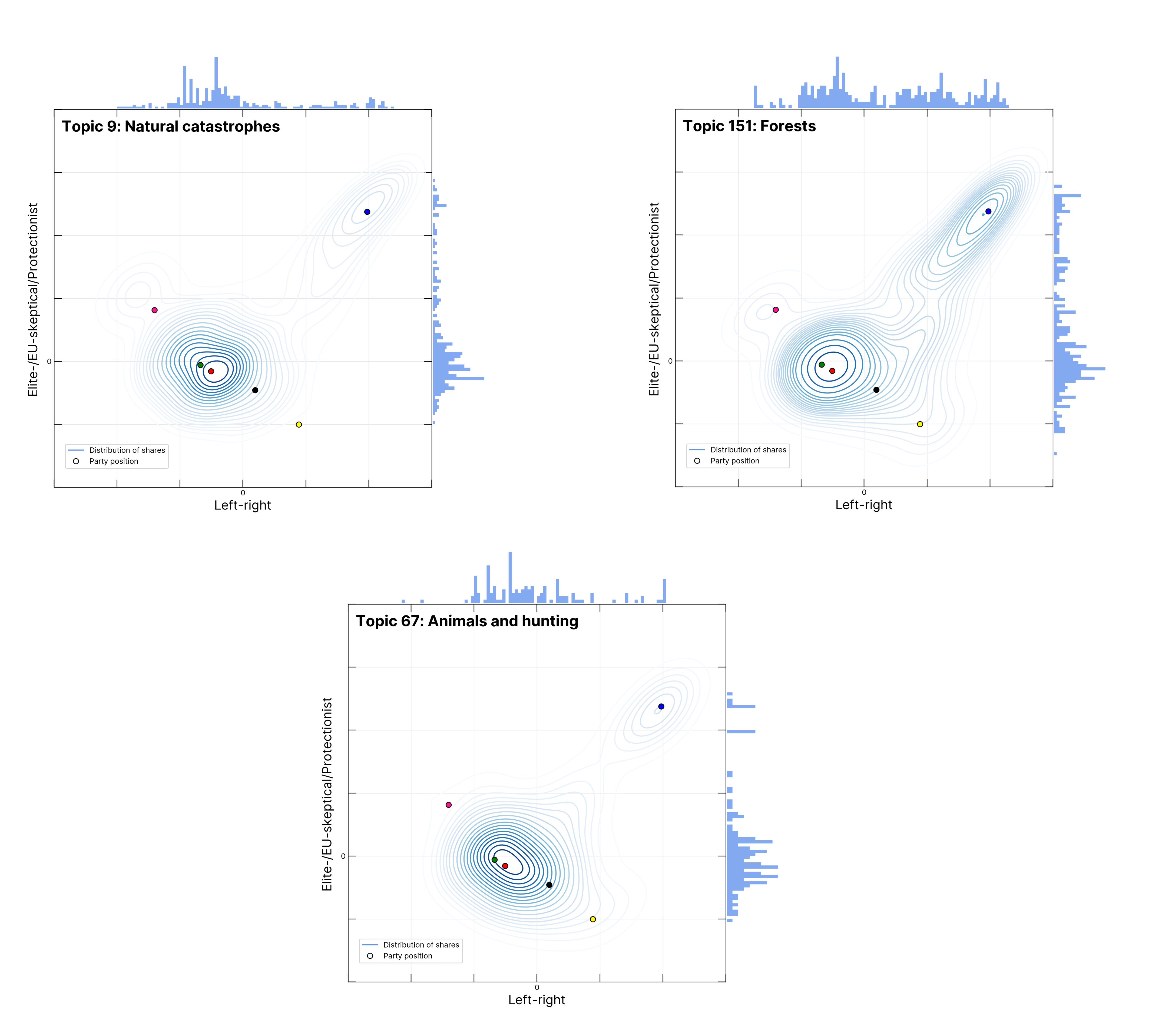}
\caption{\label{fig:other_env}Other topics attributed to metatopic `Environment and Climate Change'}
\end{figure}

See Fig. \ref{fig:other_env}.

\end{document}